\documentclass[12pt]{article}
\usepackage{graphicx,epsfig}
\input axodraw.sty
\usepackage{amsfonts,amssymb,amsmath}
\usepackage[dvips]{color}
\newcommand{\ba}{\begin{array}}
\newcommand{\ea}{\end{array}}
\newcommand{\beq}{\begin{equation}}
\newcommand{\eeq}{\end{equation}}
\def\bt{\begin{table}}
\def\et{\end{table}}
\def\bc{\begin{center}}
\def\ec{\end{center}}
\def\bi{\begin{itemize}}
\def\ei{\end{itemize}}
\def\bea{\begin{eqnarray}}
\def\eea{\end{eqnarray}}
\def\beas{\begin{eqnarray*}}
\def\eeas{\end{eqnarray*}}

\def\lp{\lambda^\prime}

\def\N0{\widetilde{\chi}^0}

\def\Cm{\widetilde{\chi}^-}
\def\Cpm{\widetilde{\chi}^\pm}

\def\snu{\tilde{\nu}}
\def\smu{\widetilde{\mu}}

\def\slep{\widetilde{\ell}}
\def\rp{R\!\!\!/ _p}

\def\slash {\!\!\!\!/}

\catcode`@=11 
\def \gsim{\mathrel{\mathpalette\@versim>}}
\def \lsim{\mathrel{\mathpalette\@versim<}}
\def \@versim#1#2{\lower0.4ex\vbox{\baselineskip\z@skip\lineskip\z@skip
     \lineskiplimit\z@\ialign{$\m@th#1\hfil##\hfil$%
     \crcr#2\crcr\sim\crcr}}} 
\catcode`@=12 

\begin{document} 
\setcounter{page}{0} 
\thispagestyle{empty} 
\begin{flushright} HIP-2010-08/TH \\
                   OSU-HEP-10-02  \end{flushright} 
\begin{center}
{\Large \sc Single production of sleptons with polarized tops at the Large Hadron Collider} 
\\ \vspace*{0.2in} 
{\large Masato Arai$^{(1)\dagger}$, Katri Huitu$^{(2)\ddagger }$,
Santosh Kumar Rai$^{(2,3)\ast}$ \\ {\rm and} Kumar Rao$^{(2)\ast\ast}$} \\
\vspace*{0.2in}
{\sl $^{(1)}$Institute of Experimental and Applied Physics, 
              Czech Technical University in Prague, \\
              Horsk\' a 3a/22, 128 00 Prague 2, Czech Republic \\
     $^{(2)}$ Department of Physics, University of Helsinki, 
              and Helsinki Institute of Physics, \\
              P.O. Box 64, FIN-00014 University of Helsinki, Finland\\
     $^{(3)}$ Department of Physics, 
              and Oklahoma Center for High Energy Physics, \\
              Oklahoma State University 
              Stillwater, OK 74078, USA \rm }$^\diamondsuit$

\vspace*{0.6in}
{\large\bf Abstract}
\end{center}

\noindent
We study the production of a single charged slepton in association
with a top quark in a R-parity violating supersymmetric model with lepton 
number violating interactions at the Large Hadron Collider.
We find that the longitudinal polarization asymmetry of the top quark      
in such a production mode is significantly different from that in              
the production of a single top or a top pair in the Standard Model 
for a wide range of slepton masses.
Our signal analysis shows that the top-slepton associate production 
leads to final states with distinct kinematic signatures, which differ
from the Standard Model background.

\vspace*{0.2in}
\vfill

\noindent
$^{\dagger}$masato.arai@utef.cvut.cz \\
\noindent
$^{\ddagger}$katri.huitu@helsinki.fi \\
\noindent 
$^{\ast}$santosh.rai@okstate.edu ({$^\diamondsuit$ current address}) \\
\noindent 
$^{\ast\ast}$kumar.rao@helsinki.fi

\section{Introduction}
The most promising candidate theory beyond the Standard Model (SM)
remains supersymmetry (SUSY), which resolves some of the 
shortcomings of the SM such as the gauge hierarchy problem.
In many supersymmetric models, a discrete multiplicative symmetry 
\cite{fayet}, R-parity,
defined by $R_p = (-1)^{3B+L+2S}$ with spin $S$, baryon number $B$,
and lepton number $L$, is often imposed on the Lagrangian
to conserve $B$ and $L$.
The definition implies that all the SM particles have $R_p = +1$, 
while all the superpartners are odd under this symmetry. 
This conservation is, however, not dictated by any fundamental 
principle such as gauge invariance or renormalizability.

The most general superpotential in SUSY, which 
respects the gauge symmetries of the SM, contains 
bilinear and trilinear terms, which do not conserve either
$B$ or $L$ and are given by
\begin{equation}
{\cal W}_{\rp} = 
 \frac{1}{2}\lambda_{ijk}  \hat{L}_i \hat{L}_j \hat{E}_k  
+\lambda^\prime_{ijk}  \hat{L}_i \hat{Q}_j \hat{D}_k
+\frac{1}{2}\lambda_{ijk}^{\prime\prime} 
 \hat{U}_{i}\hat{D}_{j}\hat{D}_{k}
+\mu_i\hat{L}_i \hat{H}_2, \label{sp} 
\end{equation}
where $\hat{L}_i,\hat{Q}_i$ are the $SU(2)$-doublet and 
$\hat{E}_i, \hat{U}_i, \hat{D}_i$ are the $SU(2)$-singlet 
superfields, respectively.
$\hat{H}_2$ is the Higgs chiral superfield. 
The indices, $i, j, k$ denote generations.
The $\lambda$, $\lambda^\prime$ and $\mu$ are the couplings of the $L$-violating interactions,
whereas $\lambda^{\prime\prime}$ are those of the $B$-violating interactions.
The co-existence of the $L$- and $B$- violating interactions leads to 
phenomenological difficulties unless R-parity violating (RPV) couplings are 
very small:
the simultaneous presence of both $L$- and $B$-violating operators 
could lead to a very rapid proton decay, especially for TeV scale sparticle 
masses.
Thus, the products of the $L$- and 
$B$-violating couplings are strictly constrained \cite{Smirnov}.
In phenomenological studies usually only one type of interaction, either
$L$- or $B$-violating, is considered. 
This can be realized, for instance, by
imposing a discrete ($Z_3$) symmetry \cite{Ibanez}.
The RPV couplings can also
lead to small neutrino masses, which are automatically generated 
either at tree- or loop-level \cite{Hall}. 
For a comprehensive review of RPV interactions, see \cite{barbier}. 
Constraints on the RPV couplings have been obtained by various 
analysis, for review see \cite{barbier,limit}.

RPV interactions can lead to new production mechanisms for single top quarks.
Associate production of single top quarks in SUSY models with RPV interactions 
have been extensively studied for several processes, see e.g. \cite{single3,single1}.
In these processes, top quarks are produced via $L$- or $B$-violating 
Yukawa type couplings given in Eq. (\ref{sp}). An important property of the top, in contrast 
to lighter quarks, is that its spin is observable since it decays before hadronization, 
owing to its extremely short lifetime.
In the SM single top quarks are produced through 
the parity violating weak interactions, leading to highly polarized top 
quarks \cite{Carlson,mahlonparke}.
The polarizations of the top quarks produced through RPV interactions
and the SM would be different, since different chiral structures are involved 
in the interaction vertices.
Effects of RPV interactions on polarized single top production and CP odd 
observables associated with top quark spin have been 
studied at a leptonic collider \cite{Chemtob}.
Polarization of top quark produced in the process 
$d\bar{d}\rightarrow t\bar{t}$ through RPV interaction has been addressed
at the Tevatron \cite{hikasa} and the Large Hadron Collider (LHC) \cite{li}. 
Even in parity conserving QCD, though the top is produced unpolarized, the spin 
of the top is correlated with that of the antitop for top pair production processes. 
Searches for new physics by using top spin correlations 
(e.g. see \cite{arai} and references therein) and single top 
polarization \cite{saurabhrohini} have been studied.

In this work,  
we study single charged slepton production in association 
with a top quark at the LHC through the interaction
$\lambda^\prime_{i3k}L_iQ_3\bar{D}_k$ in Eq.~(\ref{sp}), which
violate lepton number by one unit.
Written in terms of the component fields, the relevant terms for 
the above superpotential lead to the interaction Lagrangian
\beq
\ba{rcl}
{\cal L}_{LQ\bar{D}} &
= & \lambda^\prime_{i3k} \bigg[
   \snu_{iL}     \bar{d}_{kR}         d_{3L}
+  \widetilde{d}_{3L}    \bar{d}_{kR}         \nu_{iL}
+ (\widetilde{d}_{kR})^* \overline{(\nu_{iL})^c} d_{3L} \bigg. \\
&& \bigg. -  \slep_{iL}            \bar{d}_{kR}            u_{3L}
- \widetilde{u}_{3L}     \bar{d}_{kR}          \ell_{iL}
- (\widetilde{d}_{kR})^* \overline{(\ell_{iL})^c} u_{3L}
\bigg] + ~ \textrm{h.c}.  
\ea
     \label{gen_Lag2}
\eeq
where the superscript ``$c$" in the above equation represents the charge
conjugation of the spinor ($\psi^c=C \bar{\psi}^{T}$), $C$ being the
charge conjugation operator. 
For example $(\nu_{iL})^c\equiv (\nu^c)_{iR}$. These interactions lead to 
the production process $g d_k \rightarrow t\tilde \ell_i$, which we 
investigate in this work.
The same process has been considered in \cite{single3}, but 
the polarization of the top quark has not been investigated there.
In this work,
we investigate the effects of RPV couplings on the polarization asymmetry 
of the top quark, which is the (normalized) difference between the number 
of produced tops with spin up and spin down. 
We also show, how the properties of a polarized top are reflected in 
the decay products by performing a detailed signal analysis.
We focus on leptonic decays of the top,
which will have the cleanest signals at the LHC. 
We also highlight special kinematic variables, which are sensitive to 
polarization
effects and can lead to hints about the nature of the new physics that 
plays a role in the production of a single top. 

In Section~\ref{toppol} we discuss the top polarization and give 
the basic framework of our calculations using the spin density matrix. 
In Section~\ref{topprod} we 
present helicity amplitudes and the polarized cross sections 
for the process that we study at LHC. In Section~\ref{analysis} we show results 
for the signal and SM background analysis and present the LHC reach for the
signal in Section~\ref{lhcreach}. We summarize in Section~\ref{conclude}.

\section{Top polarization and the spin density matrix} \label{toppol}
With a large mass of $\sim 173$ GeV \cite{topmass}, the top quark has an 
extremely 
short lifetime, calculated in the SM to be $\tau_t =1/\Gamma_t \sim 5 
\times 10^{-25}$ s. This is an order of magnitude smaller than the 
hadronization time scale, which is roughly $1/\Lambda_{\rm{QCD}} \sim 
3 \times 10^{-24}$ s. Thus the top decays before it can form bound 
states with lighter quarks \cite{BDKKZ}. As a result, the spin 
information of the top, which depends on its production process, is 
reflected in characteristic angular distributions of its decay 
products. Even if the top were to form hadrons, the spin flip time 
scale induced by QCD spin-spin interactions between the top and light 
anti-quark, is of the order of $m_t/\Lambda_{\rm{QCD}}^2$, which is 
much larger than $\tau_t$. Thus the degree of polarization of an 
ensemble of top quarks can provide important information about the 
underlying physics in its production, apart from usual variables like 
cross sections. 
For a review on top quark physics and polarization 
see \cite{wbern,beneke,wagner}.
 
Top spin can be determined by the angular distribution of its decay 
products. In the SM, the dominant decay mode is $t\to b W^+$, with a 
branching ratio (BR) of 0.998, with the $W^+$ subsequently decaying 
to $\ell^+ \nu_\ell$ (semileptonic decay) or $u \bar{d}$, $c\bar{s}$ 
(hadronic decay). The angular distribution of a fermion $f$ for a top 
quark ensemble in the top rest frame has the form \cite{Kuhn}
\begin{equation}
\frac{1}{\Gamma_f}\frac{\textrm{d}\Gamma_f}{\textrm{d} \cos \theta _f}
          =   \frac{1}{2}(1+\kappa _f P_t \cos \theta _f),
\label{topdecaywidth}
\end{equation}
where $\Gamma_f$ is the partial decay width,
$\theta_f$ is the angle between the direction of the motion of 
decay fermion $f$ and the top spin vector, in the top rest frame and
\begin{equation}
 P_t=\frac{N_\uparrow - N_\downarrow}{N_\uparrow + N_\downarrow}
\label{ptdef}
\end{equation}
is the degree of polarization of the top quark ensemble, where 
$N_\uparrow$ and $N_\downarrow$ refer to the number of positive and 
negative helicity tops, respectively. 
The coefficient $\kappa_f$ is called the spin analyzing power of $f$
and it is a constant between $-1$ and $1$.
Obviously, a larger $\kappa_f$ makes $f$ a more sensitive probe 
of the top spin. 
At tree-level, the charged lepton and $d$ quark are the best spin 
analyzers with $\kappa_{\ell^+}=\kappa_{\bar{d}}=1$, while 
$\kappa_{\nu_\ell}=\kappa_{u}=-0.30$ and 
$\kappa_{b}=-\kappa_{W^+}=-0.39$ \cite{Kuhn,CJK,BNOS,MB}. Thus the 
$\ell^+$ or $d$ have the largest probability of being emitted in the 
direction of the top spin and the least probability in the direction 
opposite to the spin.
As mentioned in the introduction, at the LHC leptons can be 
measured with high precision.
Therefore, in this paper 
we focus on leptonic decays of the top quark.

Let us
consider a generic process of top production and its subsequent 
semileptonic decay $A B \to t X \to b \ell^+ \nu_\ell X$, where 
$X=P_1P_2\cdots P_{n-1}$ and $P_i(i=1,\cdots, n-1)$ are the other 
produced particles.
Since $\Gamma_t/m_t \sim 0.008$, we 
can use the narrow width approximation to write the cross section as 
a product of the $2\to n$
production cross section times the decay 
width of the top. However, in probing top polarization using angular 
distributions of the decay lepton, it is necessary to keep the top 
spin information in production and decay, thus requiring the spin 
density matrix formalism. The amplitude for $A B \to 
t(\lambda) X \to b \ell^+ \nu_\ell X$ can be written as $\sum_{\lambda} 
\mathcal{M}_P (\lambda) \mathcal{M}_D (\lambda)$ where 
$\mathcal{M}_{P,D} (\lambda)$ are the amplitudes for the production 
and decay for an on-shell top with helicity $\lambda=\pm 1$. Thus, the 
amplitude squared is of the form
\begin{eqnarray}\label{tprodk}
 |\mathcal{M}(AB \to t(\lambda) X \to b \ell^+ \nu_\ell X)|^2 
  &=&\sum_{\lambda,\lambda'}\mathcal{M}_P (\lambda) 
     \mathcal{M}_P^{*} (\lambda') \mathcal{M}_D (\lambda) \mathcal{M}_D^{*} 
    (\lambda') \nonumber \\
  &\equiv&\rho(\lambda,\lambda') \Gamma(\lambda,\lambda'),
\end{eqnarray}
where $\rho(\lambda,\lambda')=\mathcal{M}_P (\lambda) 
\mathcal{M}_P^{*} (\lambda')$ and 
$\Gamma(\lambda,\lambda')=\mathcal{M}_D (\lambda) \mathcal{M}_D^{*} 
(\lambda')$ are the $2 \times 2$ top production and decay spin 
density matrices. The off-diagonal elements encode the quantum 
mechanical interference between the production and decay amplitudes, 
which prevent a simple factorization of the process as a product of 
production and decay squared amplitudes, 
$|\sum_{\lambda}\mathcal{M}_{P}(\lambda)|^2 \,\,|\sum_{\lambda'} 
\mathcal{M}_{D}(\lambda')|^2$, where the spin information of the top 
is lost.

We consider the cross section of the $2 \to n$ production.
As in \cite{saurabhrohini}, the most general polarization density 
matrix can be parameterized as a linear combination of the Pauli 
matrices as
\begin{eqnarray}
\sigma(\lambda,\lambda^\prime)=
\frac{\sigma_{\rm{tot}}}{2}\left(
\begin{tabular}{cc}
$1+\eta_3$ & $\eta_1 - i\eta_2$ \\ 
$\eta_1 + i\eta_2$ & $1-\eta_3$
\end{tabular} \right),
\label{poldm}
\end{eqnarray}
where $\sigma(\lambda,\lambda^\prime)$ is the cross section of 
$2 \to n$ process of the top production at parton level with denoted 
spin labels and
$\sigma_{\rm{tot}}=\sigma(+,+)+\sigma(-,-)$ is the total cross section.
The (1,1) and (2,2) diagonal elements are the cross sections 
for the production of positive and negative helicity tops and 
$\eta_3$ gives the degree of {\it longitudinal} polarization
\begin{equation}
\eta_3=P_t=\frac{\sigma(+,+)-\sigma(-,-)}{\sigma(+,+)+\sigma(-,-)}.
\label{eta3def}
\end{equation}
The off-diagonal elements involving $\eta_1$ and $\eta_2$ are the 
cross sections for transversely polarized tops. The degree of transverse
polarization parallel and perpendicular to the production plane are
given by
\begin{eqnarray}
 \eta_1={\sigma(+,-)+\sigma(-,+) \over \sigma(+,+)+\sigma(-,-)},\quad
 i\eta_2={\sigma(+,-)-\sigma(-,+) \over \sigma(+,+)+\sigma(-,-)}.
\end{eqnarray}
By measuring the angular distributions of the decay lepton in the top rest 
frame (which requires reconstructing the top rest frame) analytic expressions 
for the $\eta$'s can be obtained 
by a suitable combination of lepton polar and azimuthal asymmetries 
(see \cite{saurabhrohini} for details). 
However, at a hadron machine like the LHC, reconstruction
of the top rest frame will be challenging leaving ambiguities in
the measurement of such observables.  Thus, the final state kinematics, as
discussed later, will be of utmost importance in the top events.

\section {Top-Slepton Production and Decay
} \label{topprod}
\subsection{Density Matrix for Top-Slepton Production}
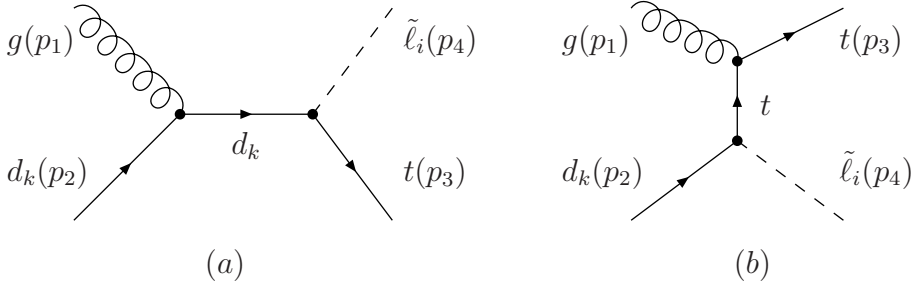
\begin{figure}[htb]
\begin{center}
\begin{picture}(800,130)(0,0)

\Gluon(100,100)(140,60){5}{5}
\ArrowLine(100,20)(140,60)
\Vertex(140,60){2}
\ArrowLine(140,60)(190,60)
\Vertex(190,60){2}
\DashLine(190,60)(220,100){5}
\ArrowLine(190,60)(220,20)
\put(75,85){$g (p_1)$}
\put(75,35){$d_k (p_2)$}
\put(225,85){$\tilde{\ell}_i (p_4)$}
\put(225,35){$t (p_3)$}
\put(160, 45){$d_k$}
\put(150,00){$(a)$}
\Gluon(310,100)(350,80){5}{4}
\ArrowLine(350,80)(390,100)
\Vertex(350,80){2}
\ArrowLine(350,50)(350,80)
\Vertex(350,50){2}
\DashLine(350,50)(390,20){5}
\ArrowLine(310,20)(350,50)
\put(285,85){$g (p_1)$}
\put(285,35){$d_k (p_2)$}
\put(390,35){$\tilde{\ell}_i (p_4)$}
\put(390,85){$t (p_3)$}
\put(360,60){$t$}
\put(350,00){$(b)$}

\end{picture}
\caption{\sl The Feynman diagrams contributing to the 
top-slepton production at the parton level at LHC via RPV 
couplings.}\label{feyngraph}
\end{center}
\end{figure}

In this subsection we derive a density matrix for 
the process of single top 
production in association with a charged slepton in R-parity 
violating SUSY. At the parton level the process is given by
\beq
g (p_1)~ d_k (p_2) \to t (p_3,\lambda_t) ~\widetilde{\ell}_i^- (p_4),
\label{prod}
\eeq
which just employs the RPV part of the Lagrangian $\left( \lambda^\prime_{i3k} 
( \slep_{iL} \bar{d}_{kR} t_{L}) \right)$ given in 
Eq.~(\ref{gen_Lag2}). 
The relevant leading order diagrams are given in Fig.~\ref{feyngraph}.
Two diagrams contribute to the production process in Eq. (\ref{prod}) with 
the down-type quark $(d_k)$ in the $s$-channel 
and the top-quark in the $t$-channel as shown in 
Fig.~\ref{feyngraph} (a) and Fig.~\ref{feyngraph} (b), respectively.

%
%

The possible set of couplings one can probe with the  
process given by Eq.~(\ref{prod}) is
\bea
\lambda^\prime_{131},\lambda^\prime_{132},\lambda^\prime_{133}\hspace*{0.5in}(\widetilde{e}~{\rm produced}), 
\nonumber \\
\lambda^\prime_{231},\lambda^\prime_{232},\lambda^\prime_{233}\hspace*{0.5in}(\widetilde{\mu}~{\rm produced}), 
\nonumber \\
\lambda^\prime_{331},\lambda^\prime_{332},\lambda^\prime_{333}\hspace*{0.5in}(\widetilde{\tau}~{\rm produced}). 
\nonumber 
\eea
Some of the above couplings are strongly suppressed and so we can 
make a case out of each such coupling that is allowed in a range, 
which gives significant cross section at the LHC. For a summary of 
the various bounds on the above couplings see 
\cite{barbier,single3}.

As discussed in the previous section, to study the polarization 
properties of the top quarks produced at LHC we need to keep the spin 
information of the top as shown in Eq.~(\ref{tprodk}). One can do that by 
writing down the helicity amplitudes for the process given by Eq.~(\ref{prod}). 
The top polarization depends on the 
choice of spin quantization axis. Choices other than the helicity 
basis have been used in which the top is almost 100$\%$ polarized 
\cite{mahlonparke}. These are useful for low velocity tops produced 
near threshold and are relevant at the Tevatron. At the LHC, since we 
expect that the tops will be highly boosted, we choose the helicity 
basis. Denoting the helicity of the gluon, top and massless down-type 
quark $d_k$ as $h$, $\lambda_t$ and $\lambda_{d_k}$ respectively, the 
non-zero $s$-channel amplitudes are
\begin{eqnarray}
\mathcal{M}_{s}(\lambda_t=+,\lambda_{d_k}=+)&=&-\sqrt{\frac{1}{2 \sqrt{s}}}\, 
g_s \, \lambda_{i3k}^{'} \left(\frac{\lambda_l}{2}\right)(1+h) 
\sqrt{E_t -p_t}\, \cos \frac{\theta}{2}, \\ 
\mathcal{M}_{s}(\lambda_t=-,\lambda_{d_k}=+)&=&\sqrt{\frac{1}{2 \sqrt{s}}}\, 
g_s \, \lambda_{i3k}^{'} \left(\frac{\lambda_l}{2}\right) (1+h) 
\sqrt{E_t +p_t}\, \sin \frac{\theta}{2},  
\end{eqnarray}
while the non-zero amplitudes for the $t$-channel are
\begin{eqnarray}
\mathcal{M}_{t}(\lambda_t=+,\lambda_{d_k}=+)&=&g_s\, \lambda_{i3k}^{'}\, 
\left(\frac{\lambda_l}{2}\right) \sqrt{\frac{\sqrt{s}}{2}}\, 
\frac{1}{(t-m_t^{2})} \left[-m_t (1+h)\sqrt{E_t +p_t} \cos \frac{\theta}{2} 
\right. \nonumber \\
&& \hskip -4.7cm \left. +(1+h) (E_t +p_t \cos \theta -\sqrt{s}) 
\sqrt{E_t -p_t} \cos \frac{\theta}{2} -(1-h) \, p_t \, \sin \theta 
\sqrt{E_t -p_t} \sin \frac{\theta}{2}\right], \\
\mathcal{M}_{t}(\lambda_t=-,\lambda_{d_k}=+)&=&g_s\, \lambda_{i3k}^{'}\, 
\left(\frac{\lambda_l}{2}\right) \sqrt{\frac{\sqrt{s}}{2}}\, 
\frac{1}{(t-m_t^{2})} \left[m_t (1+h)\sqrt{E_t -p_t} \sin \frac{\theta}{2} 
\right. \nonumber \\
&& \hskip -4.7cm \left. -(1+h) (E_t +p_t \cos \theta -\sqrt{s}) 
\sqrt{E_t +p_t} \sin \frac{\theta}{2} -(1-h) \, p_t \, \sin \theta 
\sqrt{E_t +p_t} \cos \frac{\theta}{2}\right], 
\end{eqnarray}
where $s$ and $t$ are the parton level Mandelstam variables and 
$E_t,\, p_t$ and $\theta$ are the energy, momentum and scattering 
angle of the top in the parton center-of-mass frame and $m_t, \lambda_l$ are the 
top quark mass and $SU(3)$ color matrices, while $g_s$ is the QCD 
coupling constant.

Using these helicity amplitudes the elements of the top production 
spin density matrix can be constructed. We find the following compact 
expressions for $\rho(\lambda,\lambda')$:
\bea
\rho(+,+) &=& \frac{F_1}{2} \left [ A_1 + A_2 + A_3 \cos\theta \right ], \nonumber \\
\rho(-,-) &=& \frac{F_1}{2} \left [ A_1 - A_2 - A_3 \cos\theta \right ], \nonumber \\
\rho(+,-) &=& \rho(-,+) \nonumber \\
&=& F_1 s \sqrt{s} m_t \sin\theta 
\left [ - t^2 - st + t m_{\slep}^2 + t m_t^2 - s m_{\slep}^2 - m_t^2 m_{\slep}^2 \right] 
\label{amps},
\eea
with the various functions defined by
\bea
 F_1 &=& \frac{g_s^2 ~ \lambda_{i3k}^{'2}}{24~s^2 (t-m_t^2)^2}, \nonumber  \\ 
 A_1 &=& - 2s (s+t)^2 (t-m_t^2) + 4 s t m_{\slep}^2 (s+t-m_{\slep}^2) \nonumber \\
     &&    + 2s m_t^2 (m_t^4 + 2 m_{\slep}^4 
       - 2s m_t^2 - t m_t^2 - 2 m_t^2 m_{\slep}^2), \nonumber \\ 
 A_2 &=& F_s \left[ st (s+t) - s  m_{\slep}^2 (s+3t) 
              + s m_t^2 (2s + t - 2 m_t^2 + 3 m_{\slep}^2) \right], \nonumber \\ 
 A_3 &=& s^2 t (s+t) + s m_{\slep}^2 (s^2 - t^2 - 2st + t m_{\slep}^2 - s m_{\slep}^2) \nonumber \\
     &&  + s m_t^2 (t^2 + 2s m_{\slep}^2 - t m_t^2 + m_t^2 m_{\slep}^2 - m_{\slep}^4), \nonumber \\
 F_s &=& \lambda^{1/2} \left ( s, m_{\slep}^2, m_t^2 \right ), \nonumber \\ 
 \lambda (x,y,z) &=& x^2 + y^2 + z^2 - 2xy - 2yz - 2xz.
\eea
Here $m_{\slep}$  is the slepton mass and the angular dependence 
in $t$ is given by
$$ t = m_t^2 - \frac{s + m_t^2 - m_{\slep}^2}{2} (1 - \beta_t \cos\theta),$$
where $$\beta_t = \frac{F_s}{s + m_t^2 - m_{\slep}^2}.$$
\begin{figure}[htb]
\begin{center}
\includegraphics[width=3.2in]{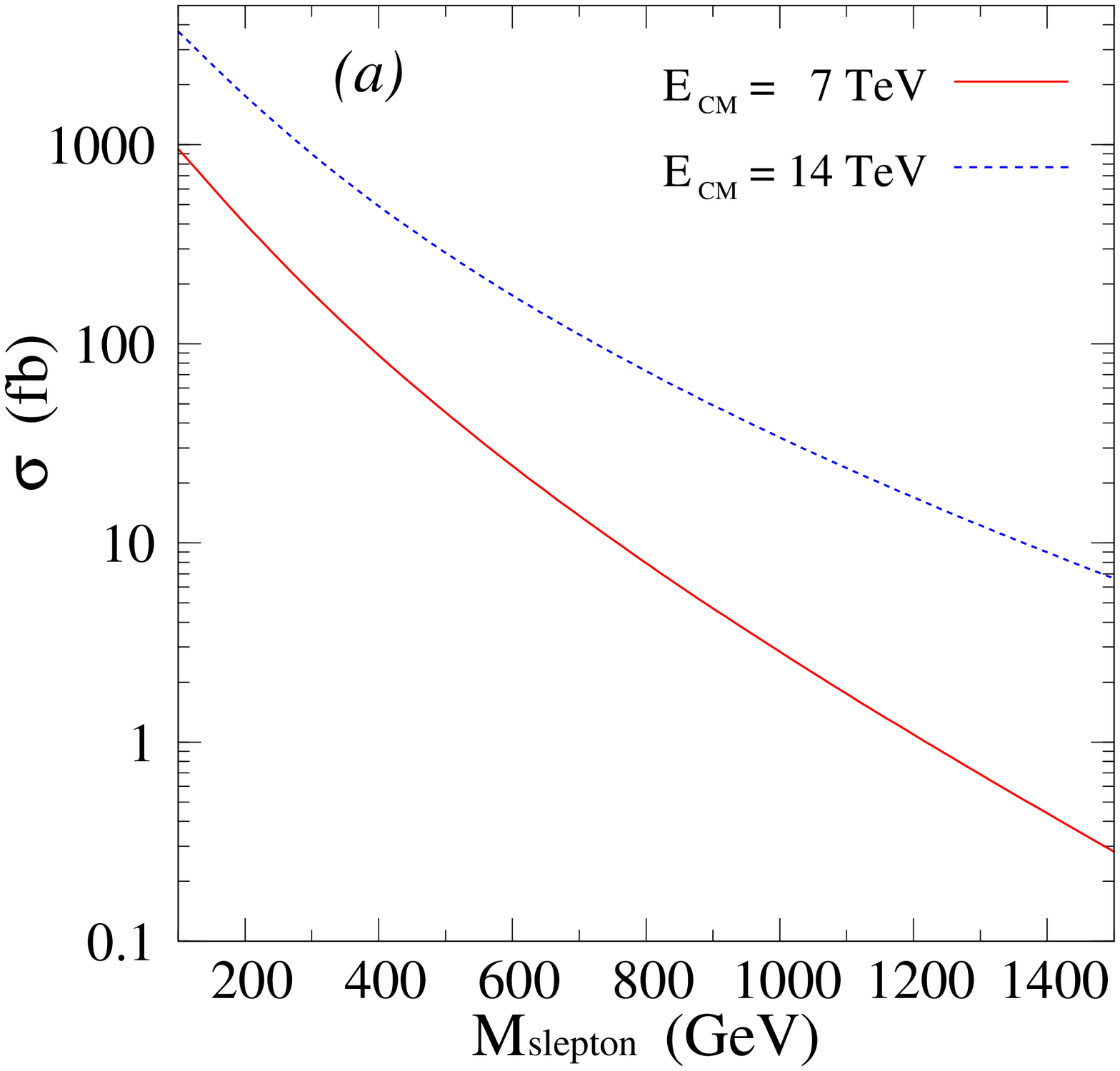} 
\includegraphics[width=3.2in]{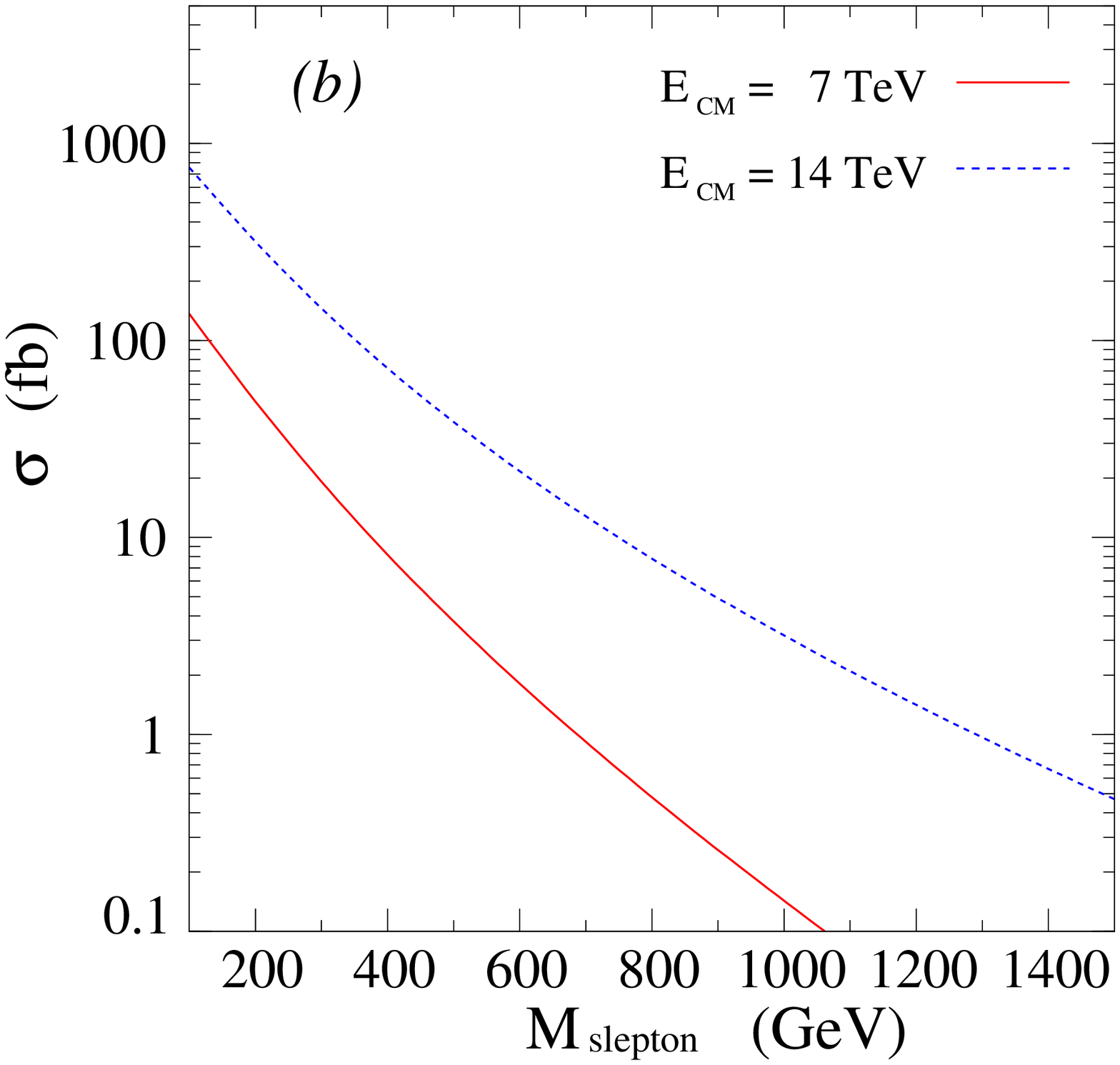} 
\caption{\sl The cross section for top-slepton production at LHC for 
two different center-of-mass energies, 7 TeV and 14 TeV. (a) shows 
the cross section for top-slepton production via $\lambda'_{i31}$ 
coupling and (b) shows the cross section for top-slepton production 
via $\lambda'_{i32}$ coupling. We choose both $\lambda' = 
0.2$.}\label{cross}
\end{center}
\end{figure}

A plot of the cross section for top-slepton production is shown in 
Fig. \ref{cross} as a function of the slepton mass, with $d$ and $s$ 
quarks in the initial state. 
Similar plots for the cross sections can also be found in \cite{single3}.
The strong constraints on the 
$\lambda^\prime_{i33} < \mathcal{O}(10^{-4})$ \cite{single3} coupling indicate
that there cannot be any significant production cross section at LHC for the 
process induced by $b$ quarks in the initial state. To illustrate the 
cross sections, we have used a fixed value of 0.2 for the 
contributing RPV couplings, which in this case are $\lambda^\prime_{i31}$ 
and $\lambda^\prime_{i32}$ for $d$ and $s$ quarks, respectively. We show the 
cross sections for two different center-of-mass energies at which the LHC 
is now planned to run, {\sl viz.} 7 TeV and 14 TeV. For the same 
strength of RPV coupling, the cross sections for the $d$ quark 
induced process (Fig.~\ref{cross}~(a)) dominates the $s$ quark induced 
process (Fig.~\ref{cross}~(b)) by nearly an order of magnitude, which is 
quite expected because of the large flux of $d$ quarks compared to 
that of $s$ quarks in the proton parton distribution. For our 
analysis, we have chosen the leading order parton density function 
(PDF) sets of CTEQ6L \cite{cteq6l} for the colliding protons.

The $(t~d_k~\slep)$ vertex is proportional to the projection operator 
$P_R=(1+\gamma_5)/2$ and thus has a chiral structure different from 
the vector and axial vector interaction vertices for $tbW$ and 
$t\bar{t}Z^0$. The latter are relevant for $tW, t\bar{b}$ and 
$t\bar{t}$ productions which are the dominant modes of 
top quark production at the LHC. We thus expect a different 
longitudinal polarization asymmetry given by Eq.~(\ref{eta3def}) for 
top-slepton production, compared to the associated $tW$ production 
via $gb\to t W$ and the top pair production processes dominated by 
the $gg,\,q\bar{q}\to t\bar{t}$ or the $W$ exchange process for 
$t\bar{b}$ production in the SM. For $tW$ production we find  
$P_t\simeq -0.25$; for $t\bar{b}$  production at LHC 
energies $P_t \simeq -0.68$, while 
$P_t \simeq \mathcal{O}(-10^{-4})$ for $t\bar{t}$ production. 
We have used the Madgraph+MadEvent \cite{madgraph} package to 
estimate these asymmetries for the SM processes. 
The very small value for the $t\bar{t}$ mode is quite expected as the 
dominant contribution comes from the gluon induced process which does 
not have any axial component in the coupling.
\begin{figure}[htb]
\begin{center}
\includegraphics[width=3.2in]{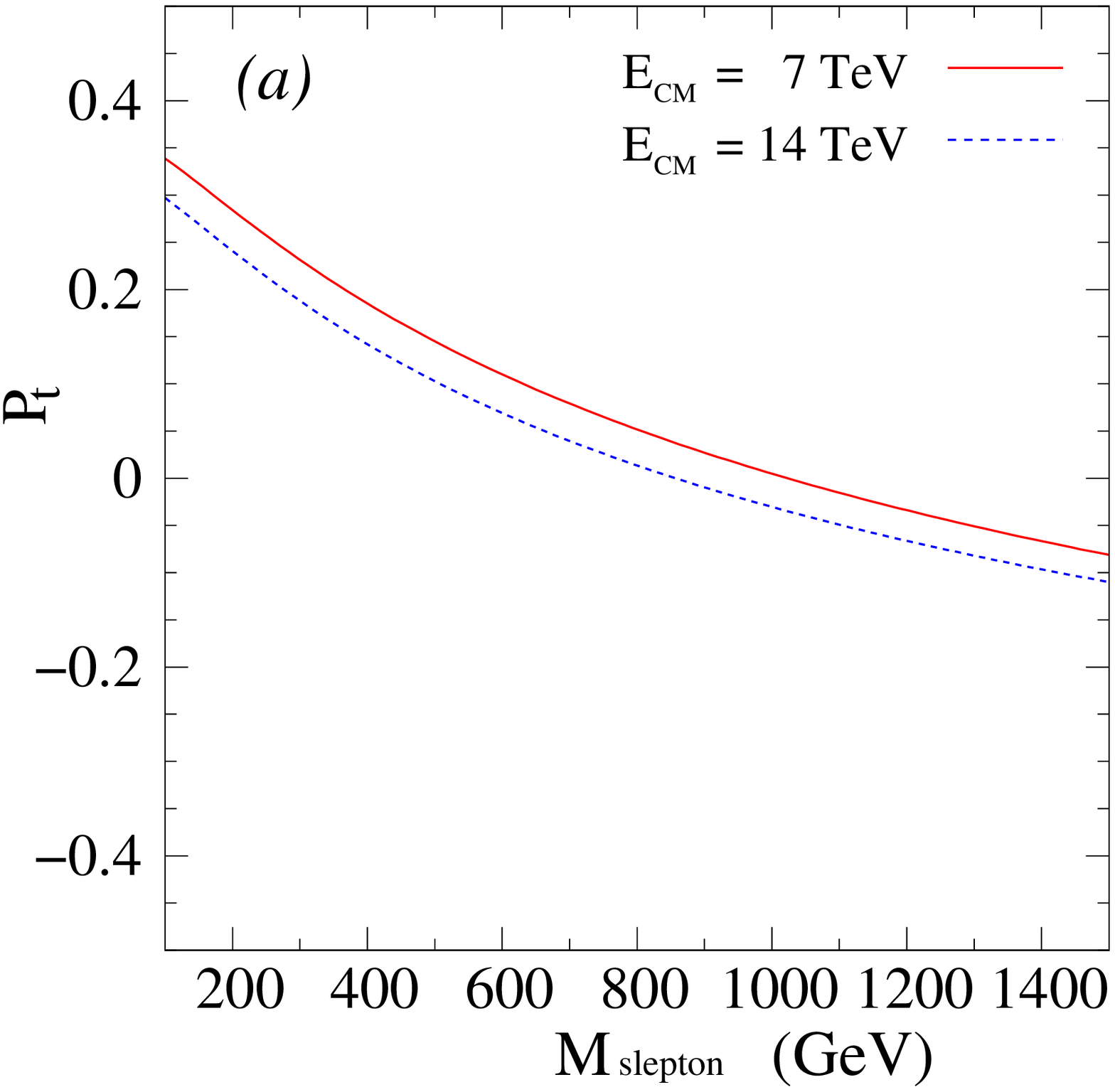} 
\includegraphics[width=3.2in]{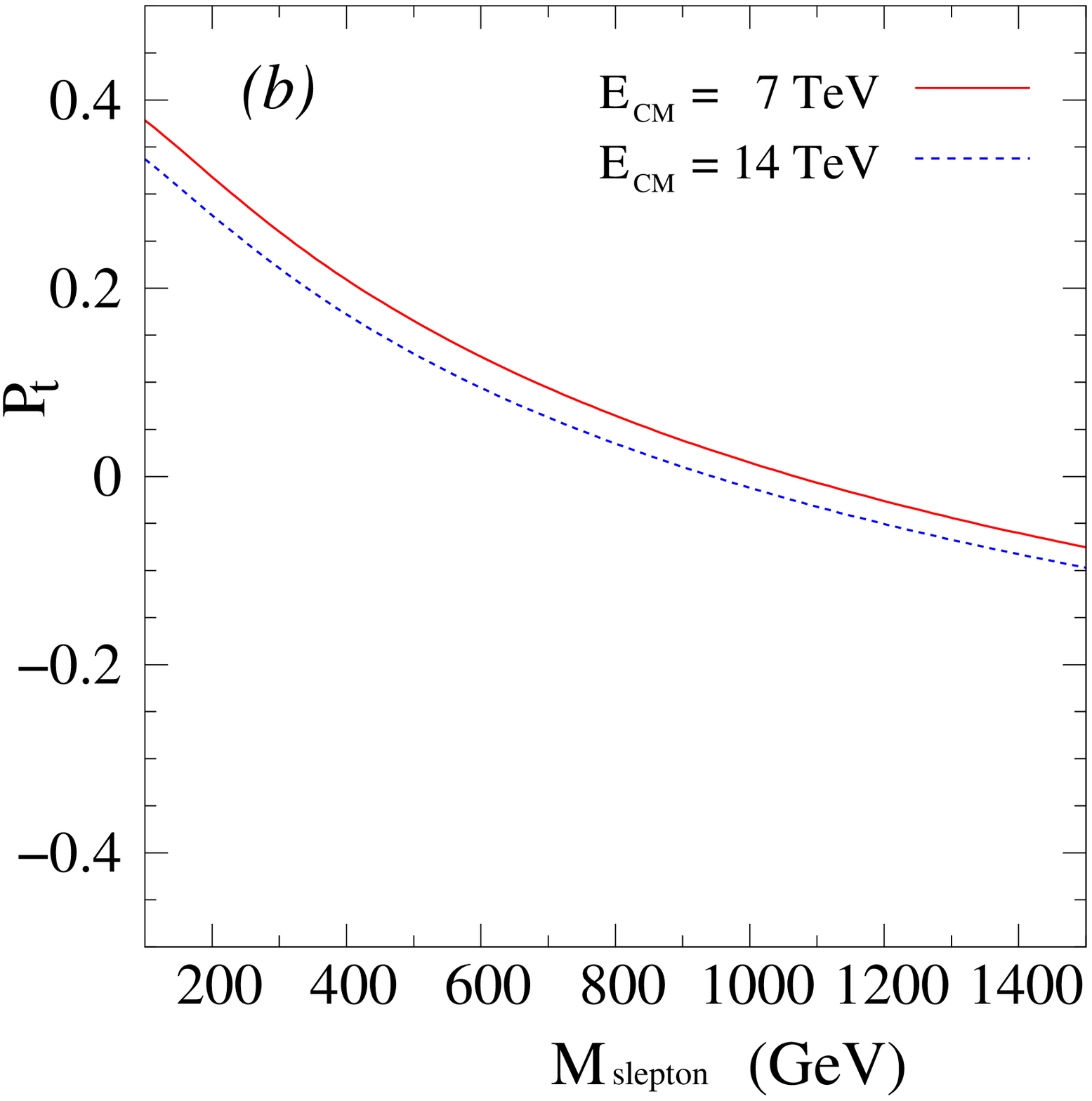} 
\caption{\sl The polarization asymmetries for top-slepton production at LHC
for two different center-of-mass energies, 7 TeV and 14 TeV.
The asymmetry $P_t$ is shown when the production is (a) via $\lambda'_{i31}$
coupling and (b) via $\lambda'_{i32}$ coupling.}\label{asymm}
\end{center}
\end{figure}
In Fig.~\ref{asymm} we plot the polarization asymmetry $P_t$ given by 
Eq.~(\ref{eta3def}) for both the $d$ and $s$ quark induced processes. 
It is worth noting that at the tree-level, $P_t$ is independent of the 
RPV coupling $\lambda_{i3k}'$ which cancels out, but is still 
sensitive to the cross sections. So although the $\lp_{i32}$ induced 
process shows slightly larger values for the asymmetry, it would be 
with limited statistics. However, the more interesting thing to note 
is that this asymmetry is significantly different from what one 
expects in the SM processes and also of opposite sign (due to the 
$P_R$ coupling), which forms the main thrust of this work. We see 
that for low values of the slepton mass, the polarization asymmetry 
can be as large as 0.33 and remains consistently positive for slepton 
masses $\leq 850$ GeV. Another interesting feature is that the top 
polarization changes sign for a slepton mass of around 870-900 GeV. 
Thus a measurement of the sign of the longitudinal polarization 
asymmetry can prove to be a useful test for 
distinguishing top-slepton production from associated top production 
modes in the SM. 
Also, such a distinct value of the polarization asymmetry
compared to the dominant mode of top quark production in the SM 
would imply that the asymmetry would leave an imprint in the 
distributions of the decay products as well as some distinct 
correlations in kinematic variables. This would also give an extra 
handle in suppressing the SM background.

\subsection{The decays of the top quark and heavy slepton}
In this section we discuss the decay of the top quark and the heavy 
slepton, needed to analyze the distinct final states, which we are 
interested in. As mentioned previously, the leptonic channel 
has the most sensitive analyzing power for the top quark 
polarization. The leptonic decay mode is further favored because of its 
cleanliness at the LHC. The respective branching ratios for top 
decays are well known, but we give the expressions for the top decay 
density matrix for completeness, which relates the decays to the 
$2\to2$ production process via Eq. (\ref{tprodk}). The top decay 
density matrix for the process $t\to b W^{+}\to b \ell^{+} \nu_\ell$ can be 
written in a Lorentz invariant form as
\beq
\Gamma(\pm,\pm)=\frac{2 g^4}{|p_W ^{2} -m_{W}^{2}+i \Gamma _{W} m_{W}|^{2}} (p_b \cdot p_\nu)
 \left[(p_\ell \cdot p_t) \mp m_t (p_\ell \cdot n_3)\right],
\eeq
for the diagonal elements and
\beq
\Gamma(\mp,\pm)=-\frac{2 g^4}{|p_W ^{2} -m_{W}^{2}+i \Gamma _{W} m_{W}|^{2}} 
\,m_t \,\,(p_b \cdot p_\nu) \,\, p_\ell \cdot (n_1 \mp i n_2),
\eeq
for the off-diagonal ones. Here the $n^{\mu}_{i}$'s ($i=1,2,3$) are the 
spin 4-vectors for the top with 4-momentum $p_t$, with the properties 
$n_i \cdot n_j =-\delta_{ij}$ and $n_i \cdot p_t =0$. 
For decay in the rest frame they take the standard form $n^{\mu}_{i}=(0, 
\delta_{i}^{k})$. 

The slepton can decay through the R-parity conserving as well
as R-parity violating modes. The specific decays it can have are
\bea
&& \widetilde{\ell}_{iL}^- \to \ell_i^- \N0_j, \nonumber \\ 
&& \widetilde{\ell}_{iL}^- \to \nu_{\ell_i} \Cm_j, \nonumber \\ 
&& \widetilde{\ell}_{iL}^- \to \bar{t} d_k \quad {\rm (via\; RPV\; coupling)}. \nonumber 
\eea

Note that for the scalar slepton, the
spin density matrix for its decay becomes trivial.
The respective partial widths for each decay mode are
given below:
\bea
 &&\Gamma(\widetilde{\ell}_{iL}^- \to \ell_i^- \N0_j) = 
 \frac{g^2 |(Z_{j2} + Z_{j1}\tan\theta_W)|^2}{32 \pi}  m_{\slep_{iL}}
 \left(1 - \frac{m_{\N0_j}^2}{m_{\slep_{iL}}^2} - \frac{m_{\ell_i}^2}{m_{\slep_{iL}}^2}\right) 
~ \lambda ^{1/2} \left( 1, \frac{m_{\N0_j}^2}{m_{\slep_{iL}}^2}, 
                           \frac{m_{\ell_i}^2}{m_{\slep_{iL}}^2} \right), \nonumber \\ \nonumber \\
 &&\Gamma(\widetilde{\ell}_{iL}^- \to \nu_{\ell_i} \Cm_j) =
 \frac{g^2 |U_{ji}|^2}{16 \pi} m_{\slep_{iL}} 
 \left( 1 - \frac{m_{\Cm_j}^2}{m_{\slep_{iL}}^2} \right )^2, \nonumber \\ \nonumber \\
 &&\Gamma(\widetilde{\ell}_{iL}^- \to \bar{t} d_k ) =
 3 \frac{\lambda_{i3k}^{'2}}{16 \pi} m_{\slep_{iL}} 
 \left(1 - \frac{m_{d_k}^2}{m_{\slep_{iL}}^2} - \frac{m_{t}^2}{m_{\slep_{iL}}^2}\right)
~ \lambda ^{1/2} \left( 1, \frac{m_{d_k}^2}{m_{\slep_{iL}}^2}, 
                           \frac{m_{t}^2}{m_{\slep_{iL}}^2} \right). 
\label{dkays}
\eea
The entries in the partial decay widths given by 
$Z_{j2}, Z_{j1}$ correspond to the elements of the mixing matrix of 
the neutralinos, while $U_{ji}$ represent the elements of the chargino 
mixing matrix. So, the respective branching ratios will depend on the 
SUSY parameters and the choice of RPV couplings. The lightest 
supersymmetric particle, which in our case is the neutralino is no 
longer stable in the RPV version of the model and will decay within 
the detector to SM particles via RPV couplings. The lightest 
neutralino has a 3-body decay \cite{Baltz:1997gd} through 
the $\lambda^\prime_{i3k}$ couplings and being a Majorana fermion gives the  
decay products
\begin{eqnarray} \label{chi0dk}
\N0_1 &\rightarrow     \nu_i b \bar{d}_k ,~~ \bar{\nu}_i \bar{b} d_k,
\end{eqnarray}
with equal probabilities. Note that we have assumed that the lightest 
neutralino is always lighter than the top quark and so its 3-body 
decay mode to a charged lepton-top quark and down quark ($\ell_i t 
d_k$) is kinematically disallowed.

\section{Signal Analysis} \label{analysis}
We now focus on the final states for our analysis and the dominant SM 
background contributing to such a final state. A quick glance at the 
decay modes suggest various possibilities to consider. As 
the slepton can decay to a lepton and a neutralino via R-parity 
conserving mode, it would be the most desirable decay mode at LHC.
To present our 
numerical results we assume a single non-zero RPV coupling given by 
$\lp_{231}$ which fixes the initial quark $d_k$ in the production 
process as the $d$ quark. Also with this choice we consider only the 
smuon $(\widetilde{\mu}_L)$ production in association with the top 
quark at LHC. We further assume that the charginos are much heavier 
than the sleptons and do not contribute in the decay of the smuon. We 
list below the possible combinations for the final states coming from 
the decay of the smuon and top quark.

\bi
\item $t$ decays to 1 $b$-jet and 2 light jets ($J$) and $\widetilde{\mu}_L$ 
decays to a $\mu^-$ and $\N0_1$.
\bi
\item  $\mu^-$ + 2$b$-jets + 3$J$ + $E\slash\,_T$.
\ei
\item $t$ decays to 1 $b$-jet and 2 light jets and $\widetilde{\mu}_L$ decays to
a $\bar{t}$ and $d$ quark.
\bi
\item 2$b$-jets + 5$J$ ($\bar{t}$ decays hadronically).
\item $\ell_j^-$ + 2$b$-jets + 2$J$ + $E\slash\,_T$ ($\bar{t}$ decays semileptonically).
\ei
\item  $t$ decays to 1 $b$-jet, $\ell_k^+$ and a neutrino while $\widetilde{\mu}_L$ decays to
a $\mu^-$ and $\N0_1$.
\bi
\item $\mu^- \ell_k^+$ + 2$b$-jets + 1$J$ + $E\slash\,_T$.
\ei
\item $t$ decays to 1 $b$-jet, $\ell_k^+$ and a neutrino while $\widetilde{\mu}_L$ decays to
a $\bar{t}$ and $d$ quark.
\bi
\item $\ell_k^+$ + 2$b$-jets + 3$J$ + $E\slash\,_T$ ($\bar{t}$ decays hadronically).
\item $\ell_j^-\ell_k^+$ + 2$b$-jets + 1$J$ + $E\slash\,_T$ ($\bar{t}$ decays semileptonically).
\ei
\ei

As pointed out earlier in Section \ref{toppol}, the effects of 
top polarization are more sensitive through the lepton in its 
semileptonic decay mode. 
In addition we hope to understand the leptonic signal at the LHC with 
much better precision compared to signals with large hadronic 
activity. Also extra efforts are in place to study the $b$-jets at LHC 
with greater efficiency. Keeping this in mind, we focus on triggering 
upon the dilepton final state with two $b$-jets and large missing 
transverse momenta with no light quark jets. We do not consider 
triggering on a light jet in the signal, since the light jet comes 
from the 3-body decay of the neutralino and is expected to be soft. 
Such a soft jet will not help against the SM background as one 
naturally expects a lot of associated soft jet multiplicity at LHC 
due to radiation. We also consider the two leptons to be of different 
flavor which makes the signal more distinct and robust. So the signal 
in question would be
\beq \label{fstate}
 pp \longrightarrow \mu^- e^+ b \bar{b} + E\slash\,_T + X.
\eeq
The most dominant SM background would come from the $t\bar{t}$ 
production as well as triple gauge boson (TGB) production ($WWZ$) 
where the $Z$ decays to $b\bar{b}$. The TGB background can be brought 
under control by a cut on the invariant mass of $b\bar{b}$. So we mainly 
focus on the $t\bar{t}$ background. It is also of interest to 
consider the $tW$ background as we would like to focus on the effect 
of top polarization on the signal and compare it with the dominant SM 
sources for a similar final state given in Eq. (\ref{fstate}). For our 
analysis we choose two different representative points in the SUSY 
parameter space given in Table \ref{susyparam}. We list only the 
relevant inputs and the masses needed for our analysis. They 
represent a light and heavy smuon which will highlight, how the 
different kinematics help in distinguishing the signal from the 
background as well as the effects of top polarization on the 
distributions.
\begin{table}[!h] 
\bc
\begin{tabular}{|c||c|c|} 
\hline {\bf Parameters} & {\bf I} & {\bf II} \\ \hline
\hline $(M_1,~M_2)$ & $(100,~300)$ & $(100,~500)$ \\ 
\hline $A_i$        & $-1000$ & $-1500$ \\ 
\hline $(\mu,~\tan\beta)$& $(250,~10)$ & $(600,~5)$ \\ 
\hline $(M_{\ell L},~M_{\ell R})$ & $(200,~200)$ & $(500,~500)$ \\ \hline 
\hline $(M_{\N0_1},~M_{\Cpm_1})$  & $(93,~218)$ &  $(97,~478)$ \\ 
\hline $(m_{\slep L},~m_{\slep R})$& $(205,~205)$ & $(502,~502)$ \\ 
\hline $m_{\snu L}$  & $190$  & $496$ \\ 
\hline $\lp_{231}$  & $0.2$  & $0.5$ \\
\hline 
\end{tabular}
\ec
\caption{\sl Representative points in the MSSM parameter space and 
the relevant mass spectrum used in the analysis. All mass parameters 
are given in units of GeV. To generate the mass spectrum for the 
supersymmetric particles we have used the code Suspect 
\cite{Djouadi:2002ze}.}
\label{susyparam}
\end{table}
To calculate and generate events for the final state given by 
Eq. (\ref{fstate}) and study the effects of top polarization on the 
kinematics, we are required to keep the spin information of the top 
quark in its production and decay. For this purpose we have used the
package {\sl Madgraph+MadEvent} \cite{madgraph} with its explicit use 
of helicity amplitudes. We have included the 
relevant vertices for the RPV interactions and used 
this package to calculate the signal as well as the SM background, 
namely the final states coming from the $t\bar{t}$ and $tW$ 
production. We must point out here that the $tW$ background is only 
considered for the purpose of comparing the distributions of the top 
decay products to highlight the polarization effect due to different 
interaction vertices involved in its production.

For triggering on the final states, we set the following kinematic cuts
\bi
\item The charged leptons must have a minimum $p_T$ of 10 GeV and lie 
within the rapidity gap given by $\vert \eta^\ell \vert < 2.5$.
\item The $b$-jets in the final state must satisfy $p_T>20$ GeV and 
respect the rapidity cut of $\vert \eta^b \vert < 2.5$.
\item The final states must account for a minimum missing transverse 
energy, $E\slash\,_T > 50$ GeV.
\item To ensure proper spatial resolution between the final state 
particles we demand that $\Delta R_{\ell_i\ell_j} > 0.2, \Delta 
R_{\ell b} > 0.4$ and $\Delta R_{bb} > 0.7$ where $\ell$ represents 
the charged leptons. The $\Delta R$ between two particles is defined 
as $\Delta R_{AB} = \sqrt{\Delta \phi_{AB}^2 + \Delta\eta_{AB}^2}$, 
with $\Delta\phi$ and $\Delta\eta$ being the separation in the 
azimuthal angle and the rapidity of the two particles.
\ei
With this set of kinematic cuts we calculate the signal for the two 
representative points given in Table~\ref{susyparam}. We have assumed 
a $b$-jet identification efficiency of 50\%. We must point out here
that for our parton-level analysis, the $b$-jet is represented by the 
parent $b$-quark produced in the final state. Since the signal is sensitive to 
the strength of the RPV coupling, whose limits are dependent on the 
squark masses, we use conservative values for the couplings by setting
the squark masses to 1 TeV. For our choice of the RPV couplings listed in 
Table~\ref{susyparam}, we find that the signal satisfying the above
set of kinematic cuts at LHC with the initial 
run of $\sqrt{s}=7$ TeV, is 4.3 fb for $m_{\smu_L}=205$ GeV, while it is 
8.4 fb for $m_{\smu_L}=502$ GeV. The main decay modes that contribute to the 
signal depend on the respective branching ratios. The branching 
ratios for the two cases are given by
\bea
 BR (\smu_L \to \mu \N0_1) = 0.76,~~BR (\smu_L \to \bar{t} d) = 0.24~~(m_{\smu_L}=205~GeV),  \nonumber \\ 
 BR (\smu_L \to \mu \N0_1) = 0.09,~~BR (\smu_L \to \bar{t} d) = 0.90~~~(m_{\smu_L}=502~GeV). \nonumber 
\eea
For the lighter smuon, the signal is completely dominated by the 
contributions coming from the R-parity conserving decay of smuon. 
Although the RPV decay is around 24\%, the small branching fraction 
for the top quark decay to the leptonic mode makes its contribution 
very small. However, for the heavier smuon the RPV decay is 90\% and 
hence contributes significantly to the final states. In fact we find 
that it contributes to about 58\% of the signal. For the 14 TeV 
collisions at LHC with the same kinematic cuts on the events, there 
is a significant increase in the signal, 
which is around 13.2 fb for $m_{\smu_L}=205$ GeV while it is 
48 fb for $m_{\smu_L}=502$ GeV. In comparison to the signals, we find 
that the SM background for the same kinematic selection cuts, is very 
large with the $t\bar{t}$ contributions at leading order (LO) coming out 
to be $\sim 145$ fb for the 7 TeV machine while it is $\sim 780$ fb for the 14 TeV 
collisions. The K-factors for both the signal and background will 
help in reducing the uncertainties in the statistics. However even 
with the LO results of the signal, one can see that with a high 
luminosity of 100 fb${}^{-1}$, the signal can be significant.
\begin{figure}[t!] 
\begin{center}
\includegraphics[height=3.2in,width=3.2in]{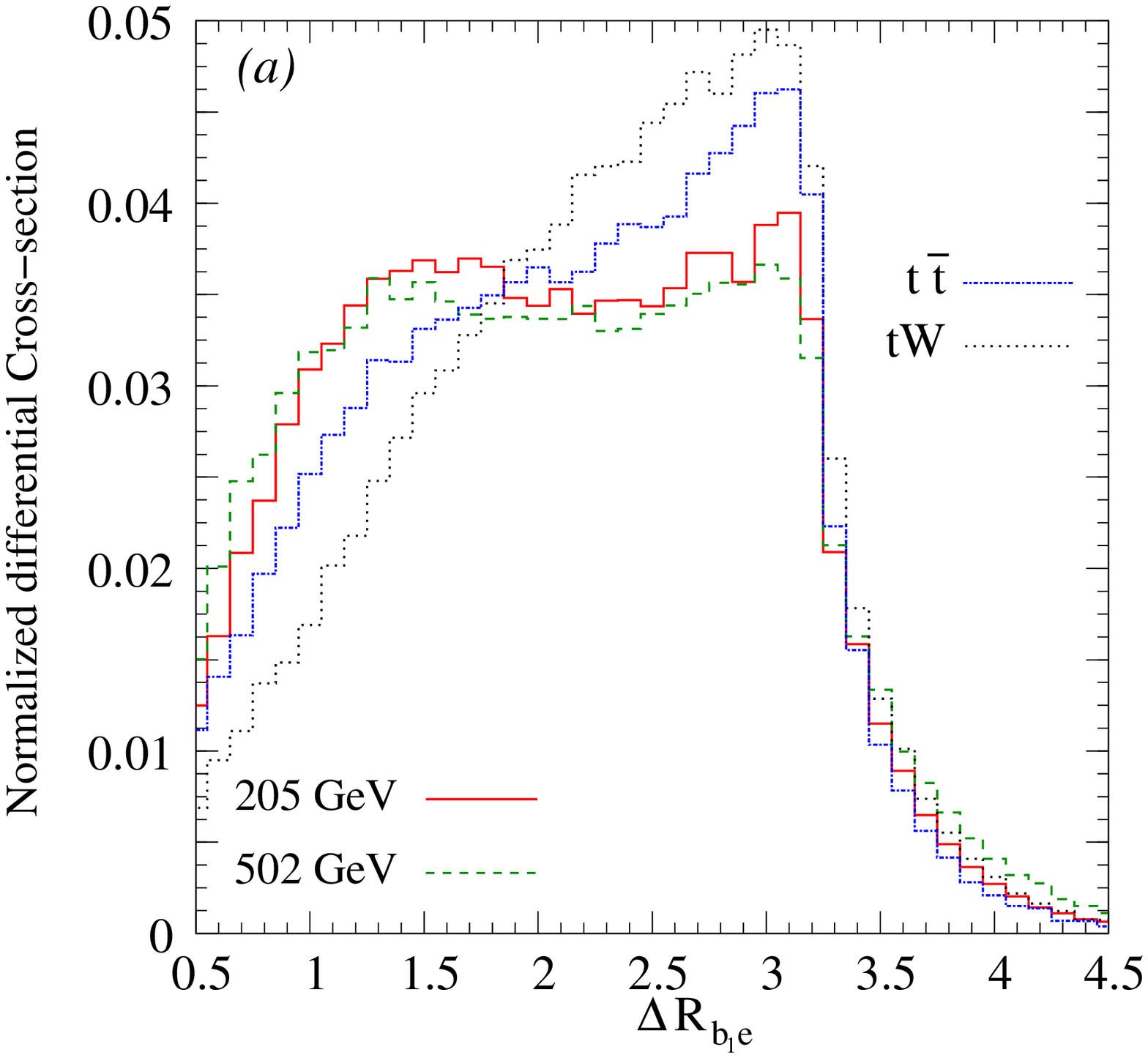}
\includegraphics[height=3.3in,width=3.2in]{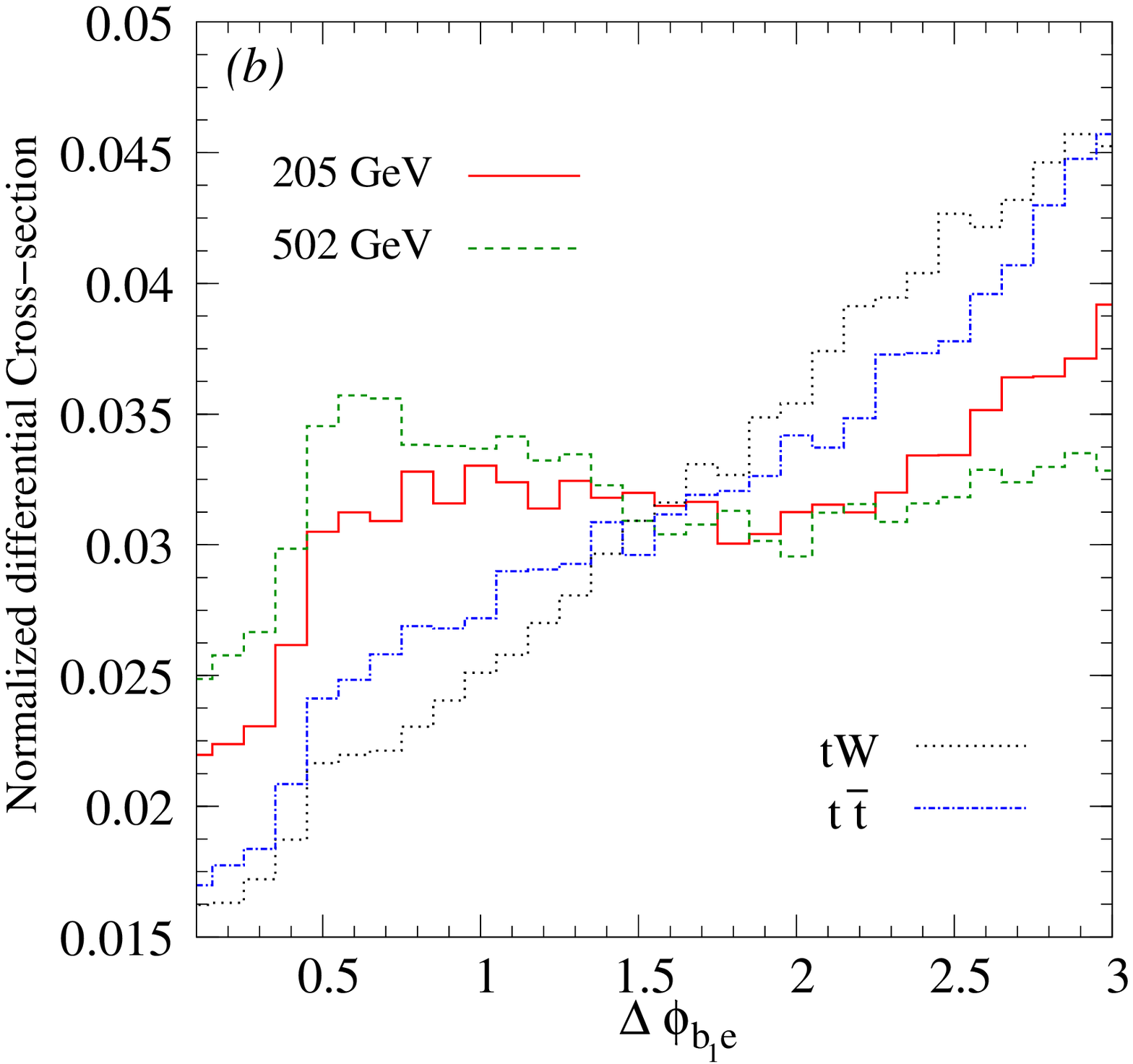}
\caption{\sl Showing the normalized distributions for the 
signal with two different smuon mass of 205 GeV and 502 GeV as well as the
competing SM background for associated top quark production. In
(a) we show the separation in $\Delta R_{b_1 e}$ while in (b) we show
the azimuthal separation $\Delta\phi_{b_1 e}$. Here both $e$ and $b_1$ 
which represents the leading $b$-jet, primarily come from the top quark 
decay.}\label{phiplots}
\end{center}
\end{figure}

It is also worth noting that as the muon in the final state comes 
from a heavy smuon, it will have a larger $p_T$ compared to the 
muon which comes from the decays of $W$ boson in the  
SM. So a stronger $p_T$ cut for the muon in the final state will also 
help in reducing the SM background \cite{single3}. As an 
estimate we put a stronger cut on the muon by choosing $p_T^\mu > 40$ 
GeV at the $\sqrt{s}=7$ TeV collisions. The signal becomes 3.2 fb and 
6.4 fb for the light and heavy smuon, respectively. But there is a much
stronger suppression for the SM background coming from $t\bar{t}$, 
which becomes 80 fb at the LO. This increases the significance of 
the signal a lot. In fact a much stronger $p_T$ cut on the muon 
is desirable for the heavy smuon signal which would be very effective 
in reducing the large SM background \cite{single3}. We discuss more on the 
effects of various kinematic cuts on the signal and background at the LHC 
in Section \ref{lhcreach}.

Let us now try and see what effect the strong polarization 
asymmetry (Fig. \ref{asymm}) for the associated top production at LHC 
via the chirality violating coupling $\lp_{231}$ has on the final 
states. In Fig. \ref{phiplots} we show the distribution of particles 
which come directly from the decay of the top quark. In 
Fig.~\ref{phiplots} (a) we plot the spatial separation $(\Delta R)$ in 
the $(\eta,\phi)$ plane between the leading $b$-jet $(b_1)$ and the 
electron, coming primarily from the top quark decay, for the signal
and the SM background. The normalized 
distributions show that there is a significant difference in the 
distributions for the decay products of the top quark for the signal 
when compared with the SM subprocesses. The difference is much better 
highlighted in the $\Delta \phi$ distribution as shown in 
Fig.~\ref{phiplots} (b). This shows that a $\Delta\phi$ difference is a 
clear highlight of the top polarization effect on the distributions. Although 
it does not give a direct estimate of the polarization of the top quark, it 
gives a very clear indication of its importance as a probe to study the effect of 
a chirality violating coupling responsible for its production when compared to SM. 
We find that the above distributions do not change much for muon $p_T$ cuts 
of less than 40-50 GeV. However much stronger $p_T$ cuts on the muon result in events 
with highly boosted tops which significantly affect the angular correlation.
\begin{figure}[ht] 
\begin{center}
\includegraphics[height=3.3in,width=3.2in]{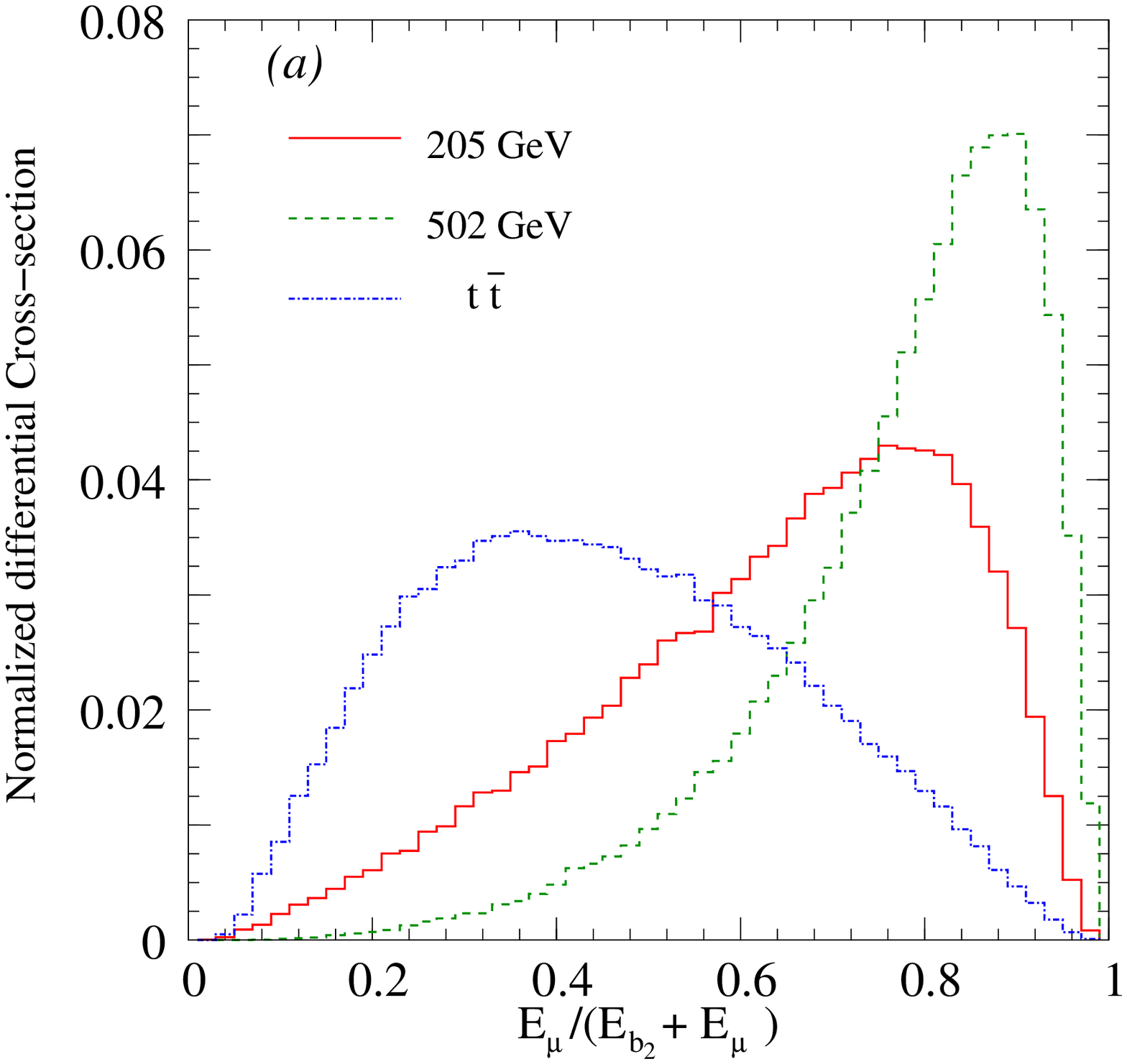}
\includegraphics[height=3.3in,width=3.2in]{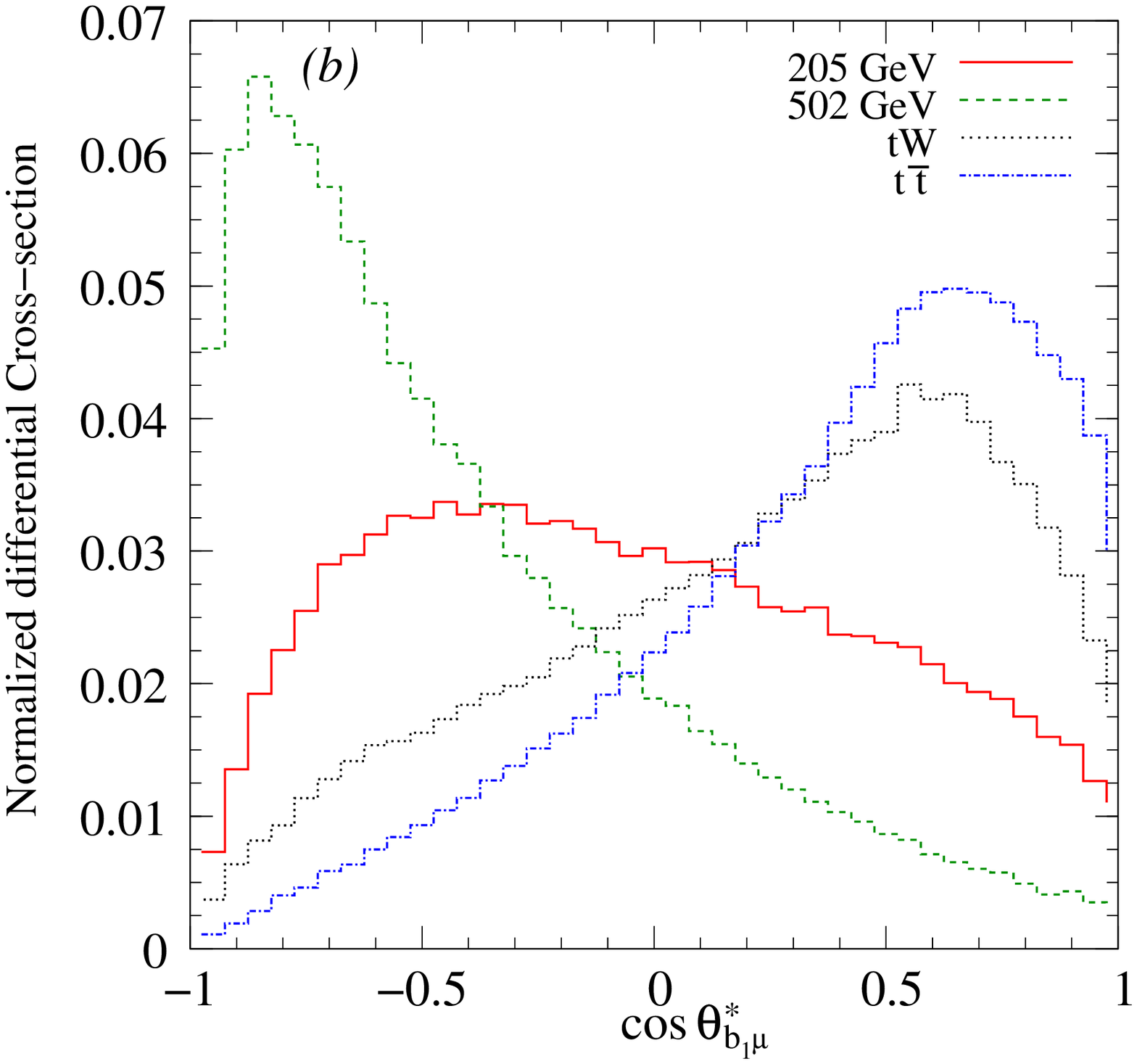}
\caption{\sl Showing the normalized distributions for the 
signal with two different smuon mass of 205 GeV and 502 GeV as well as the
competing SM background for associated top quark production. In
(a) we show the relative strength of the energy of $\mu$ and in (b) we show
the angular variation of the $\mu$ in a special frame (explained in text).}\label{eplots}
\end{center}
\end{figure}

In Fig. \ref{eplots} we show a few kinematic distributions, which are 
sensitive to the muon energy as well as to the nature of the mother 
particle it comes from. In Fig.~\ref{eplots}(a) we plot the normalized 
cross section with respect to the ratio between the energy of the 
muon and the sum of the energy of the muon and the sub-leading $b$-jet, 
$E_\mu/(E_{b_2}+E_\mu)$. This is a variable which directly reflects 
the energy strength of the muon in the smuon decay chain. As the 
sub-leading $b$-jet dominantly comes from the 3-body decay of the 
neutralino, this ratio will always peak for values greater than 0.5 
as the muon comes from the primary decay of smuon and carries energy 
depending on the mass difference between the smuon and neutralino. 
The SM background contribution to this ratio is peaked for values 
less than 0.5 since the muon in this case comes from the decay of the 
$W$ boson, while the sub-leading $b$-jet always comes from the top 
decay. The ratio, however, does depend on the relative mass differences 
as is evident from Fig.~\ref{eplots} (a) which shows the sharp shift in 
the value for the heavier smuon as compared to the lighter smuon. The 
ratio will also be sensitive to the $p_T$ cuts. Nevertheless, it is 
an effective variable to distinguish the signal from the SM 
background.

Fig.~\ref{eplots} (b) shows an even more interesting distribution. It is 
the cosine of the angle constructed in the rest frame of the muon and 
the leading $b$-jet and represents the angle between the direction of 
the muon in this frame with respect to the boost direction of the 
muon plus $b$-jet system. We find that this variable is quite sensitive 
to the nature of the mother particle of the muon. As shown in the 
figure, both for the $t\bar{t}$ and the $tW$ backgrounds the muon, 
which always comes from the decay of the $W$ boson, is peaked for 
$\cos\theta_{b_1\mu}^* > 0$ while it peaks for 
$\cos\theta_{b_1\mu}^* < 0$ for the case when it comes from the decay 
of the scalar particle ($\smu_L$). This gives a very clear hint of the 
different nature of the spin of the particle, if not the spin itself.  
It will be interesting to see if this variable can help to 
distinguish between signals of universal extra dimensions model, 
which is often called the "bosonic" SUSY. One could also consider various 
asymmetries in the above variables, which can also prove to be useful 
tools in distinguishing the signal from the SM background.

We have till now focused only on the R-parity conserving decays of 
the smuon to highlight the signal. However, the muon can also come 
from the semileptonic decay of the top quark, if the
\begin{figure}[htb] 
\begin{center}
\includegraphics[height=3.2in,width=3.2in]{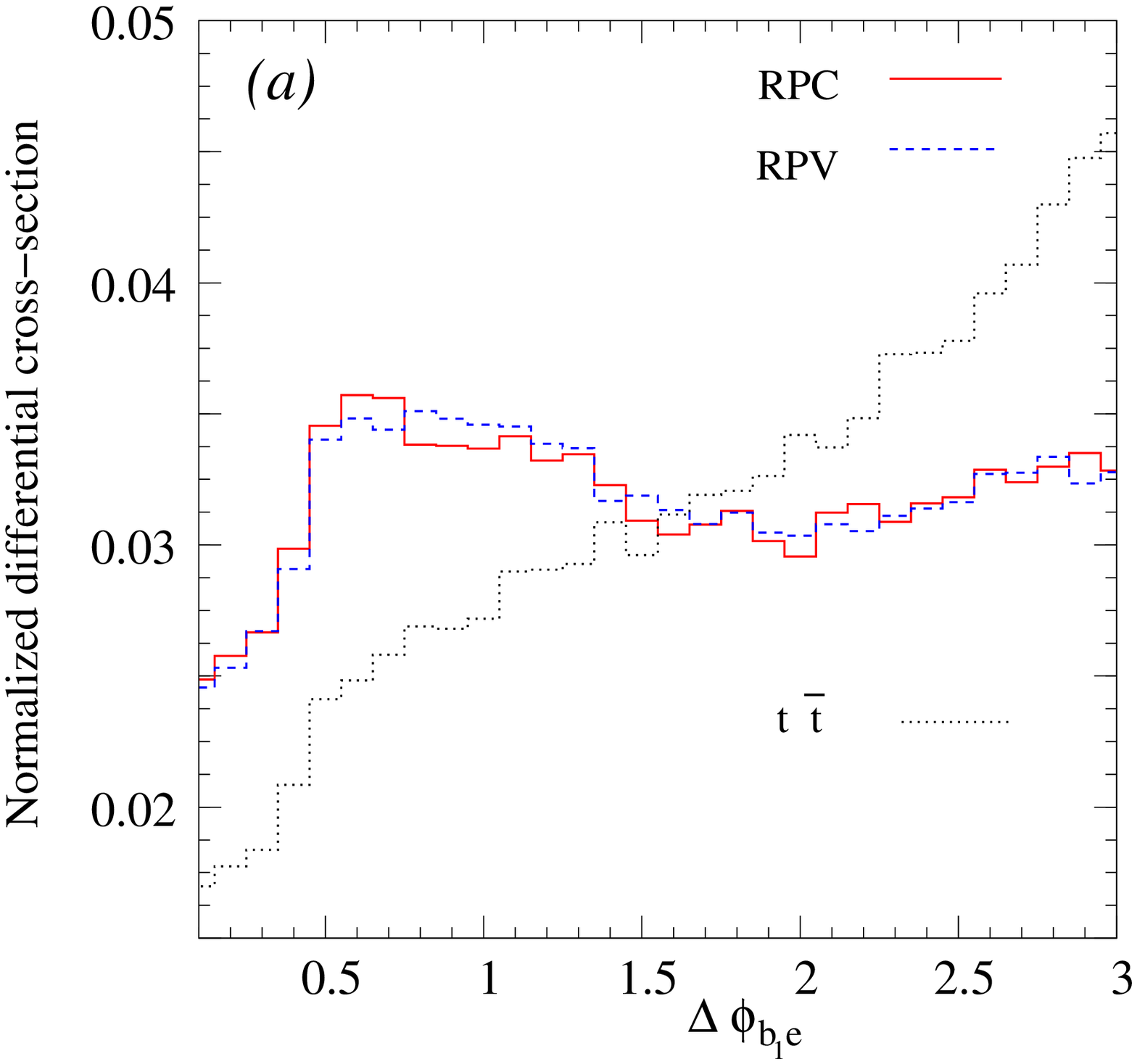}
\includegraphics[height=3.3in,width=3.2in]{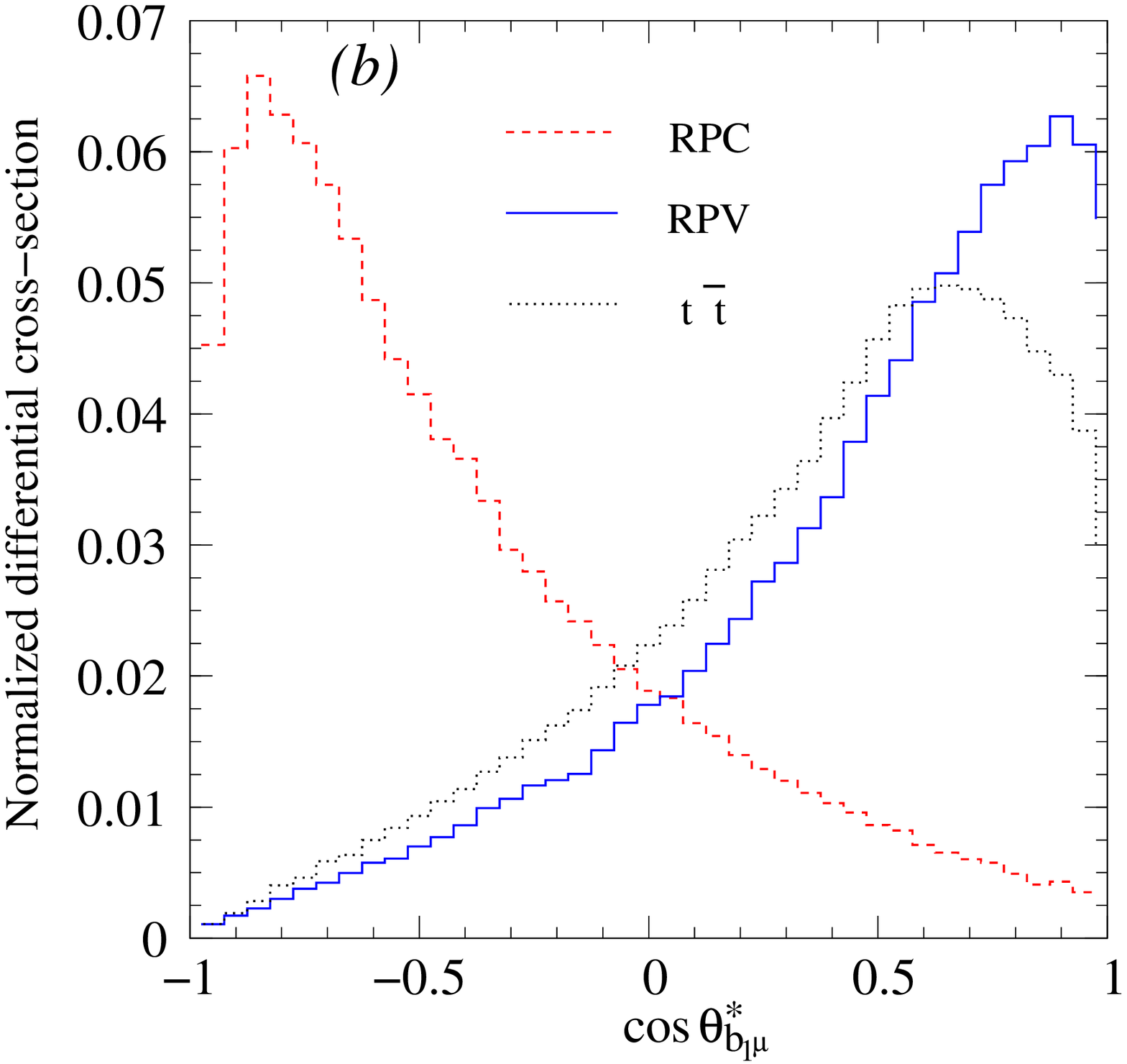}
\caption{\sl Showing the normalized distributions of the signal for 
the R-parity conserving (RPC) and R-parity violating (RPV) decays of 
the smuon of mass 502 GeV as well as the SM background from 
$t\bar{t}$. In (a) we show the azimuthal separation $\Delta\phi_{b_1 
e}$ (Fig. \ref{phiplots} (b)). In (b) we show the angular variation of 
the $\mu$ (Fig. \ref{eplots} (b)).}\label{mplots}
\end{center}
\end{figure}
RPV decay dominates as is the case for 
$m_{\smu_L}=502$ GeV. We now focus on how this may affect the 
distributions. As already stated before, the distributions shown 
in Fig. \ref{phiplots} represent the decay products of
the top quark and are, therefore, not affected as long as the top quark 
is produced in association with the slepton. To show this, we plot the normalized 
distributions again for $m_{\smu_L}=502$ GeV in Fig.~\ref{mplots} (a) 
where we consider both the R-parity conserving decay given by 
$\smu_L^-\to\mu^- \N0_1$ and the RPV decay 
$\smu_L\to\bar{t} d$ separately. Note that the muon ($\mu^-$)
now comes from the decay of $\bar{t}$ and should therefore also highlight
the polarization of the antitop. We show that the $\Delta\phi_{b_1e}$ 
distribution remains unaffected and also the normalized $\Delta\phi_{b_1\mu}$ 
distribution is found to overlap with the $\Delta\phi_{b_1e}$ distribution. Thus 
both the muon as well as the electron carry the information of the chirality
violating vertex of the $t$ and $\bar{t}$ with the $\smu_L$. The other plot 
(Fig.~\ref{eplots} (b)) which represents the spin information 
of the mother particle should however change since the muon now comes 
from the semileptonic decay of the anti-top quark and should be 
similar to the $t\bar{t}$ contribution. This, in fact turns out to be 
same when considered separately as shown in Fig. \ref{mplots}(b). This 
clearly shows that the angular distribution shown is directly 
sensitive to the nature of the mother particle it originates from.

\section{Signal potential at the LHC} \label{lhcreach}
In this section we discuss the signal potential for the single slepton production at
the LHC. As the dominant background comes from the $t\bar{t}$ production, we would like   
\begin{table}[!t]
\begin{center}
\begin{tabular}{|l||l|} \hline
\textsl{cut-1}  & $p_T^\mu > 10$ GeV, $p_T^e > 10$ GeV, $p_T^b > 20$ GeV, $E\slash\,_T > 50$ GeV, \\ 
                & $|\eta^{\ell,b}|<2.5$, $\Delta R_{e\mu} > 0.2$, $\Delta R_{\ell b} > 0.4$, 
                 $\Delta R_{bb} > 0.7$  \\ \hline 
\textsl{cut-2a} & $p_T^\mu > 60$ GeV, $p_T^e > 10$ GeV, $p_T^b > 20$ GeV, $E\slash\,_T > 50$ GeV, \\ 
                & $|\eta^{\ell,b}|<2.5$, $\Delta R_{e\mu} > 0.2$, $\Delta R_{\ell b} > 0.4$, 
                 $\Delta R_{bb} > 0.7$ \\ \hline 
\textsl{cut-2b} & $p_T^\mu > 100$ GeV, $p_T^e > 10$ GeV, $p_T^b > 20$ GeV, $E\slash\,_T > 50$ GeV, \\ 
                & $|\eta^{\ell,b}|<2.5$, $\Delta R_{e\mu} > 0.2$, $\Delta R_{\ell b} > 0.4$, 
                 $\Delta R_{bb} > 0.7$ \\ \hline
\textsl{cut-3}  & $p_T^\mu > 10$ GeV, $p_T^e > 10$ GeV, $p_T^b > 20$ GeV, $E\slash\,_T > 50$ GeV, \\ 
                & $|\eta^{\ell,b}|<2.5$, $\Delta R_{e\mu} > 0.2$, $\Delta R_{\ell b} > 0.4$, 
                 $\Delta R_{bb} > 0.7$, $\Delta\phi_{b_1e} < 1.5$ \\ \hline
\textsl{cut-4a} & $p_T^\mu > 60$ GeV, $p_T^e > 10$ GeV, $p_T^b > 20$ GeV, $E\slash\,_T > 50$ GeV, \\ 
                & $|\eta^{\ell,b}|<2.5$, $\Delta R_{e\mu} > 0.2$, $\Delta R_{\ell b} > 0.4$, 
                 $\Delta R_{bb} > 0.7$,  $\Delta\phi_{b_1e} < 1.5$ \\ \hline 
\textsl{cut-4b} & $p_T^\mu > 100$ GeV, $p_T^e > 10$ GeV, $p_T^b > 20$ GeV, $E\slash\,_T > 50$ GeV, \\ 
                & $|\eta^{\ell,b}|<2.5$, $\Delta R_{e\mu} > 0.2$, $\Delta R_{\ell b} > 0.4$, 
                 $\Delta R_{bb} > 0.7$, $\Delta\phi_{b_1e} < 1.5$  \\ \hline
\end{tabular}
\caption{\textsl{Different choices for kinematic cuts on the final states 
$\mu^- e^+ b \bar{b} + E\slash\,_T + X$ to study the LHC reach.}} \label{cuts}
\end{center}
\end{table}
to see which cuts would be relevant for suppressing the background without affecting
much of the signal cross section. The most important kinematic variable turns out to
be the muon transverse momentum. The muon coming from the primary decay of the smuon
has a large $p_T$ as compared to the muon coming from the semileptonic decay of the 
top quark. Also most of the kinematic variables described in the previous section
depend on our cut on the muon $p_T$. So our choice of the transverse
momenta cut on the muon also becomes quite relevant for studying the top polarization
effects, since the angular correlations (in the decay products) are likely to get 
washed away for very boosted top quarks.

In Table \ref{cuts}, we list different set of kinematic cuts with changes in the cuts for
the muon $p_T$ and the $\Delta \phi_{b_1e}$ variable and show at what significance the signal can be 
observed at LHC. The {\sl cut-1} corresponds to the minimal set where we have the distinct
correlation in the azimuthal angular distributions highlighting the top polarization effects
as shown in Fig. \ref{phiplots}. The {\sl cut-2a} and {\sl cut-2b} represent strong $p_T$ cuts
on the muon of 60 GeV and 100 GeV respectively, while {{\sl cut-3} corresponds to a 
$\Delta \phi_{b_1e} < 1.5$ cut (to exploit the large asymmetry seen for SM in Fig. \ref{phiplots}(b)
for {\sl cut-1}) to reduce the SM background. The cuts defined by {\sl cut-4a} and {\sl cut-4b} 
again represent strong $p_T$ cuts on the muon of 60 GeV and 100 GeV respectively, with the 
additional cut of $\Delta \phi_{b_1e} < 1.5$.

\begin{table}[!t]
\begin{center}
\begin{tabular}{|l|c|c|c||c|c|c|} \hline
&\multicolumn{3}{|c||}{$\sqrt{s}=7$ TeV}&\multicolumn{3}{|c|}{$\sqrt{s}=14$ TeV} \\
\cline{2-4}\cline{5-7}
  \multicolumn{1}{|c|}{Cuts} &$m=205$ GeV & $m=502$ GeV & SM
&$m=205$ GeV & $m=502$ GeV & SM \\ \hline
\textsl{cut-1}  & 17.2 & 33.6 & 579.3 & 52.8 & 192.3 & 3127.5 \\ \hline
\textsl{cut-2a} &  9.8 & 19.8 & 182.2 & 35.5 & 111.7 & 1026.7 \\ \hline
\textsl{cut-2b} &  --  &  --  &   --  & 17.5 &  84.7 &  334.2 \\ \hline
\textsl{cut-3}  &  6.8 & 16.1 & 218.6 & 24.5 &  93.9 & 1192.0 \\ \hline
\textsl{cut-4a} &  4.6 &  9.5 &  79.6 & 17.5 &  55.2 &  455.1 \\ \hline
\textsl{cut-4b} &   -- &  --  &  --   &  9.6 &  42.5 &  177.6 \\ \hline  
\end{tabular}
\caption{\textsl{The leading order cross sections (in {\it fb}) for the 
kinematic cuts listed in Table~\ref{cuts} on the final states 
$\mu^- e^+ b \bar{b} + E\slash\,_T + X$ for the signal and
the SM background. No efficiency factors included for the $b$-jets in this table.}} \label{csec_cuts}
\end{center}
\end{table}
%
In Table \ref{csec_cuts} we give the total cross section for the final states
$\mu^- e^+ b \bar{b} + E\slash\,_T + X$ for the signal for two values of the
smuon mass and also the SM background at LHC with the different kinematic cuts listed in
Table \ref{cuts}. We can see that the strong cuts on the $p_T$ of muon turn out to be most 
effective in improving the significance of the signal. We have not included any $b$-tagging
efficiency factors for the cross sections given in Table \ref{csec_cuts}. Including 
a $b$-tagging efficiency of 50\% one can find that for the case of smuon of
mass 205 GeV and with $\sqrt{s}=7~(14)$ TeV center-of-mass energy at LHC, one can
get a significance $S=3.63~(5.54)$ with {\sl cut-2a} for a luminosity of 100 $fb^{-1}$. 
It is however worth noting that with {\sl cut-1} one still has appreciable signal 
significance ($S=3.57~(4.72)$) for the lighter smuon. We have defined the {\it significance} 
as $S=\dfrac{N_s}{\sqrt{N_{SM}}}$, where $N_s$ represents the number of events coming from the 
RPV contribution and $N_{SM}$ is the number of events for the SM background. We have excluded 
the values for the cross sections for the 7 TeV run at LHC for the more stronger cuts 
given by \textsl{cut-2b} and \textsl{cut-4b} in Table \ref{csec_cuts} which also have 
strong suppressions for the signal.
\begin{figure}[!htb] 
\begin{center}
\includegraphics[height=3.0in,width=3.2in]{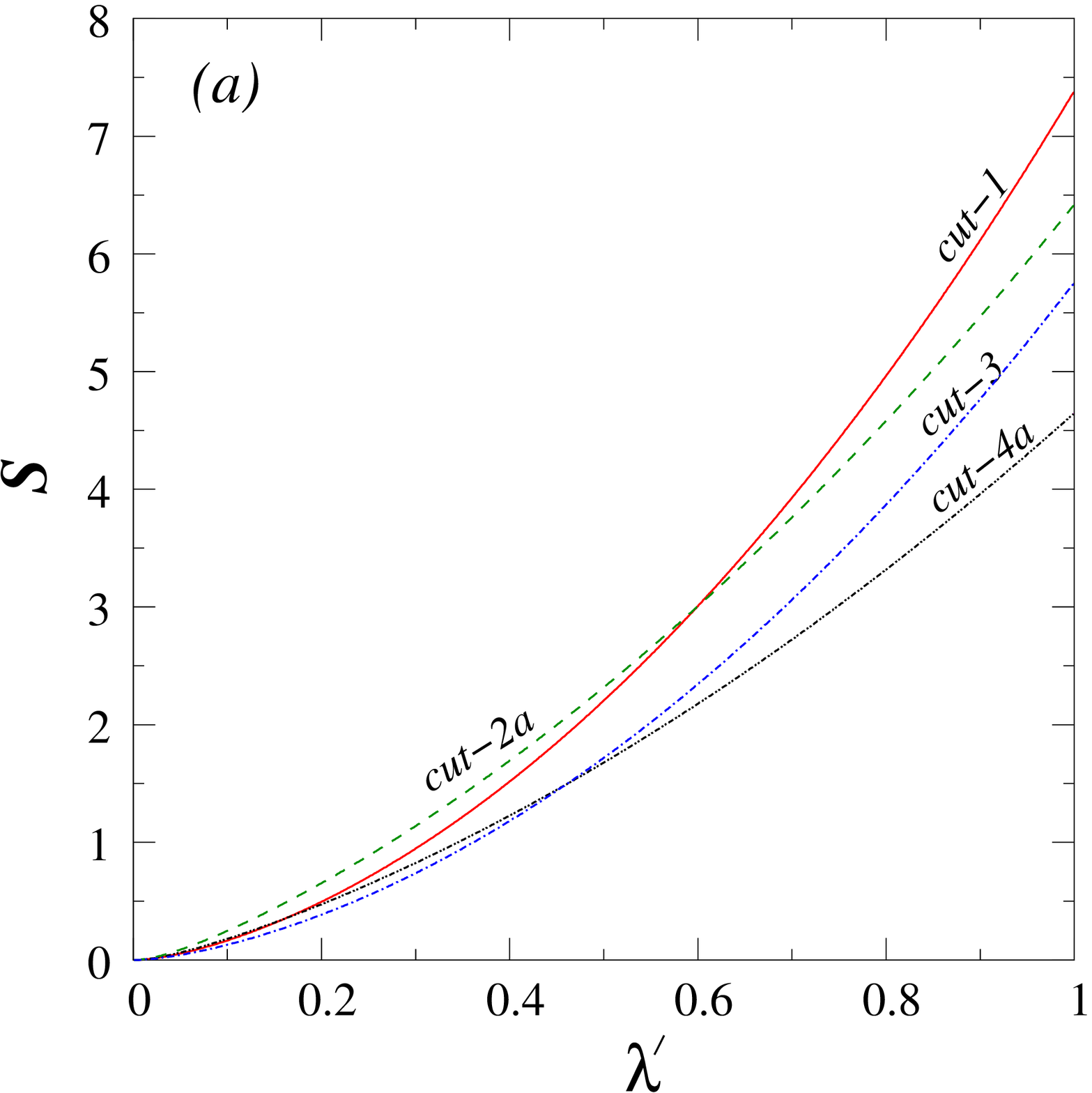}
\includegraphics[height=3.0in,width=3.2in]{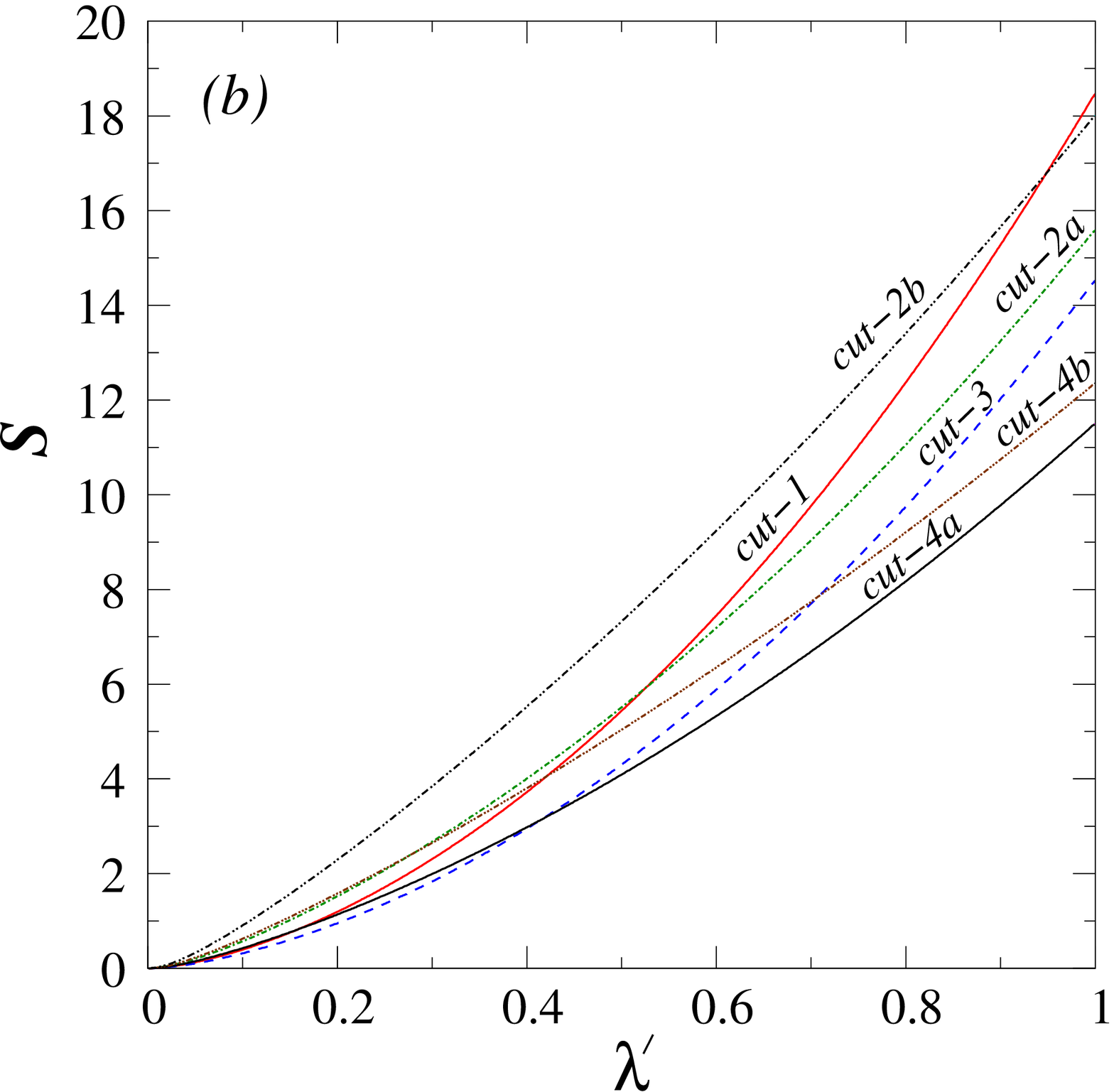}
\caption{\sl The significance ($S$) is shown for the different kinematic cuts, 
as a function of RPV coupling ($\lambda^\prime$) for a fixed integrated luminosity ($L=10 ~fb^{-1}$) 
and smuon mass ($m_{\smu_L}=502$ GeV) for two different center-of-mass energies, 
(a) $\sqrt{s}=7$ TeV and (b) $\sqrt{s}=14$ TeV.}\label{plot502}
\end{center}
\end{figure}

In Fig. \ref{plot502} we plot the significance as a function of the RPV coupling
$\lambda^\prime$ for the smuon of mass, $m_{\smu_L}=502$ GeV. It is important
to note here that both the production cross section for the single top production
with the smuon and the decay properties of the smuon depend on the value of 
$\lambda^\prime$. For small values of $\lambda^\prime$ the R-parity conserving
decay of the smuon ($\smu_L^-\to\mu^- \N0_1$) is the primary source for the 
signal. However, as the RPV coupling becomes larger, the RPV decay mode of the smuon 
($\smu_L\to\bar{t} d$) becomes the dominant source for the signal and 
the muon then mostly comes from the semileptonic decay of the anti-top. This would mean 
that a strong $p_T$ cut on the muon also causes suppression of the signal for large 
$\lambda^\prime$ coupling. This is illustrated in Fig. \ref{plot502} where we have chosen 
the integrated luminosity as 10 $fb^{-1}$. We find that the significance for 
the strong cut on the muon $p_T>100$ GeV ({\sl cut-2b}) becomes comparable 
to {\sl cut-1} for larger values of $\lambda^\prime$ as compared to smaller 
values of $\lambda^\prime$.
It is also found that the cut on the variable $\Delta \phi_{b_1e}$ ({\sl cut-3}) 
also gives a reasonably high significance, but is not as effective as the cut on 
the muon $p_T$.    

\section{Summary}  \label{conclude}
We have studied the single production of a slepton in association 
with a top quark at the LHC. Our analysis has focused on describing 
the effects of the top polarization on the particular signal of an 
associated charged slepton and we have shown that the Lorentz 
structure at the production vertex for the top can lead to very 
distinct signals, which have not been considered in the literature. We 
find that the polarization asymmetry is significantly different from 
the SM expectation for a very wide range of slepton mass accessible 
at the LHC. However, the work relies on violating  
R-parity in SUSY so that we can produce a single slepton. 
A natural extension to this work would be to look at the associated 
production of top with charged Higgs \cite{beccaria, wip} which would 
be challenging as the leptonic mode (first 2 generations) of decay is suppressed. 
However, as seen in Fig.~\ref{mplots} (a), the top polarization effects would 
still show up in some kinematic distributions. Another interesting 
variable that we have found through our analysis is the 
$\cos\theta^*$ variable for a final state particle.  It is found to be 
sensitive to the spin of the mother particle it originates from. It 
would be interesting to study this variable in order to distinguish 
models, which predict particles with different spins.

In our numerical analysis we have chosen only one non-zero 
$\lp$ coupling. This study can also be replicated for other $\lp$ 
couplings, leading to different final states. However, the 
interesting kinematic features studied and highlighted for the final states 
would still hold and can prove to be useful tools in constraining the 
RPV couplings. 

\vspace{0.3in}
{\large\bf Acknowledgments:}\\
S.K.R. would like to thank A. Khanov and F. Rizatdinova for useful 
discussions.
This work was supported in part by the Research Program MSM6840770029 
and by the project of International Cooperation ATLAS-CERN 
of the Ministry of Education, Youth and Sports of 
the Czech Republic (M.~A.). 
K.H. and K.R. gratefully acknowledge the support from the Academy of Finland
(Project No.~115032).
S.K.R. is supported by US Department of Energy, 
Grant Number DE-FG02-04ER41306.


\end{document}